\documentclass[rmp,aps,nofootinbib,reprint,endfloats,showpacs]{revtex4-1}
\pdfoutput=1
\usepackage{hyperref}
\usepackage{graphicx}
\usepackage{epsfig}
\usepackage{amsmath}

\def\wig#1{\mathrel{\hbox{\hbox to 0pt{%
          \lower.5ex\hbox{$\sim$}\hss}\raise.4ex\hbox{$#1$}}}}

\def\Dwa{$\,$\uppercase\expandafter{\romannumeral5}$\,$}

\def\sles{\lower2pt\hbox{$\buildrel {\scriptstyle <}
   \over {\scriptstyle\sim}$}}
\def\sgreat{\lower2pt\hbox{$\buildrel {\scriptstyle >}
   \over {\scriptstyle\sim}$}}
\def\sharpnull#1{}

\begin{document}

\title{Perspectives on Core-Collapse Supernova Theory}
\date{\today}

\author{Adam Burrows}
\affiliation{Department of Astrophysical Sciences\\ Princeton University\\ Princeton, NJ 08544 USA
\\ e-mail: burrows@astro.princeton.edu\\ URL: \url{http://www.astro.princeton.edu/~burrows}}


\setcounter{page}{1}

\begin{abstract}
Core-collapse theory brings together many facets of high-energy and nuclear astrophysics
and the numerical arts to present theorists with one of the most important, yet frustrating,
astronomical questions: ``What is the mechanism of core-collapse supernova explosions?" A review of all 
the physics and the fifty-year history involved would soon bury the reader in minutiae that could easily
obscure the essential elements of the phenomenon, as we understand it today. Moreover, much 
remains to be discovered and explained, and a complicated review of an unresolved subject in flux
could grow stale fast.  Therefore, in this paper I describe what I think are various important 
facts and perspectives that may have escaped the attention of those interested in this puzzle.
Furthermore, I attempt to describe the modern theory's physical underpinnings and briefly summarize 
the current state of play. In the process, I identify a few myths (as I see them) that
have crept into modern discourse. However, there is much more to do and humility in the face of this age-old challenge 
is clearly the most prudent stance as we seek its eventual resolution.
\end{abstract}



\maketitle

\tableofcontents
\twocolumngrid

\section{Introduction}
\label{intro}

Stars are born, they live, and they die.  Some, the most massive ($\gtrsim 8 {\rm M}_{\odot}$), die
explosively, spawning in the process neutron stars or ``stellar-mass" black holes while
littering the interstellar medium with many of the elements of existence.
But what is this process by which a star's multi-million year life is terminated abruptly
and violently within seconds, then announced over months via the brilliant optical
display that is a supernova explosion?  Fifty years of theory, calculation, and observation
have not definitively answered that question, though a vivid picture of the mechanism
and terminal scenario of the dense core of a massive star is emerging.

The solution to the puzzle of the mechanism of core-collapse supernova (CCSN) explosions
involves more than just obtaining simulation explosions on supercomputers.  If this weren't the case,
theorists would have solved this poser many times (and have!).  Rather, the ``solution" 
involves quantitatively explaining a host of astronomical facts that surround the supernova
phenomenon.  These include, but are not limited to: 1) the canonical explosion energy
of $\sim$10$^{51}$ ergs (defined as one ``Bethe"), along with its putative distribution 
from $\sim$0.1 Bethes to $\sim$10 Bethes.  Explosion energy is potentially a function of progenitor mass,
rotation rate, magnetic fields, and metallicity.  To date, no one has come close to achieving this central goal; 
2) the residual neutron star mass and its distribution as a function of star.
This involves more than simply noting that the Chandrasekhar mass (M$_{Ch}$) of $\sim$1.4 M$_{\odot}$
is similar to the gravitational masses of well-measured neutron stars (though this
fact is relevant to zeroth-order). The proto-neutron star (PNS) fattens by accretion during 
the respite before explosion, so the density and angular-momentum profiles in the progenitor 
core, the time of explosion, and the amount of fallback are all determining factors that 
are intimately tied to the mechanism and its unfolding. The branch map connecting progenitor to 
either neutron star or stellar-mass black hole final states is a related goal; 3) the nucleosynthetic
yields as a function of stellar progenitor. Which stars yield how much of which elements is a combined function
of a) the pre-explosion stellar evolution to the ``onion-skin" structure of progressively 
heavier elements as one tunnels in to the central core and b) the explosion itself, which
determines the mass cut and the degree of explosive nucleosynthetic reprocessing.  The
yields of the elements between calcium and the iron peak (inclusive) are particularly sensitive
to the explosion process, constituting as they do the inner ejecta; 4) the high average 
pulsar proper motion speeds. Radio pulsars are the fastest population of stars in the galaxy, 
with average speeds of $\sim$350 km s$^{-1}$, but ranging beyond $\sim$1000 km s$^{-1}$. 
Asymmetries in the explosion itself and simple momentum recoil are natural culprits, 
but we don't have an explanation for their observed speed spectrum, nor for which progenitors 
give birth to the low proper-motion subclass of neutron stars bound as accreting X-ray sources 
to globular clusters; and 5) supernova explosion morphologies and ejecta element spatial 
distributions. Instabilities and asphericities in the explosion itself are compounded by
instabilities during the propagation of the supernova shock wave through the progenitor
star and the circumstellar medium to create a debris field that is anything but spherical
and neatly nested. Even a qualitative identification of the signatures of the explosion
process itself in the density and element distributions of the expanding supernova blast
and subsequent supernova remnant (SNR) would be an advance.

The literature on core-collapse supernova theory is vast and entails a fifty-year history of  
hydrodynamics and shock physics, radiative transfer, nuclear physics (at many junctures),
neutrino physics, particle physics, statistical physics and thermodynamics, gravitational physics,
and convection theory. The explosion signatures, along with those listed in the previous paragraph, 
are photon light curves and spectra, neutrino bursts, gravitational wave bursts, and 
meteoritic and solar-system isotope ratios. Thousands of researchers have at one time or another 
been engaged, many as careerists. To attempt to summarize or synthesize this literature, even with a
focus on the theory of the mechanism, would be a gargantuan undertaking.  Moreover, since the 
fundamental mechanism has not been satisfactorily and quantitatively demonstrated, such an
ambitious review might seem premature.   

Nevertheless, there have in the past been attempts to review core-collapse explosion theory, and some of these
contain useful information and perspectives.  All, however, due to the inexorable evolution
of the subject as researchers have struggled towards the ultimate goal of understanding, and due
to the vastness of the task, are of limited scope and clearly trapped in time. This is natural. However, for those readers who 
desire a Cook's tour in the tradition of a standard, though helpful, review of the various 
aspects of core-collapse theory, I list here a sampling. Neither the samples, nor the sampling
are complete, and many of these papers were not written as reviews. For overall 
perspectives, I point to Bethe (1988;1990), Janka (2001,2012), Janka et al. (2007), 
Burrows et al. (1995,2007ab), Herant et al. (1994), Kotake, Sato, \& {Takahashi (2006), 
Kotake et al. (2012ab), Burrows (2000), and Mezzacappa (2005).  For neutrino microphysics, one can consult 
Dicus (1972), Bruenn (1985), Burrows, Reddy, \& Thompson (2006), Tubbs \& Schramm (1975), Freedman, Schramm, \& Tubbs (1977),
and Langanke \& Martinez-Pinedo (2003). For equation of state issues, there are Lamb et al. (1981),
Lattimer (1981), and Lattimer \& Swesty (1991). For massive star evolution and nucleosynthesis, 
good sources are Burbidge et al. (1957), Weaver, Zimmerman, \& Woosley (1978), Woosley \& Weaver 
(1986,1995), Thielemann, Nomoto, \& Hashimoto (1996), Weaver \& Woosley (1993), Nomoto et al. 
(1997), Woosley, Heger, \& Weaver (2002), and Heger, Woosley, \& Spruit (2005). For the connection with 
gamma-ray bursts, one can turn to Woosley \& Bloom (2006) and for gravitational wave signatures
there are Ott (2009), Kotake (2012), M\"uller et al. (2004), and 
M{\"u}ller} {et~al.} (2012).  There are many papers on the computational issues specific
to core collapse, but the papers by Arnett (1966,1967,1977), Imshennik \& Nad{\"e}zhin (1973), 
Bowers \& Wilson (1982), Bruenn (1985), Mayle \& Wilson (1988), Liebend\"orfer et al. (2001), 
Liebend\"orfer, Rampp, Janka, \& Mezzacappa (2005), Burrows et al. (2000), Livne et al. (2004,2007), and Swesty \& Myra (2009) 
are collectively educational. Ph.D. theses by R. Mayle, A. Marek, and B. M\"uller are particularly 
informative.  Classics in the subject, though not particularly comprehensive, 
include Burbidge et al. (1957), Colgate \& Johnson (1960), Hoyle \& Fowler (1960,1964), Fowler \& Hoyle (1964), Colgate \& 
White (1966), Arnett (1966), LeBlanc \& Wilson (1970), Wilson (1971), Mazurek (1974), Sato (1975), 
Bethe et al. (1979), and Bethe \& Wilson (1985). 

To this set of ``reviews," and in particular to the list of ``classics," one could add many 
others. However, anyone well-versed in the papers on the above short list will be well-informed
on most of supernova theory, if not completely {\it au courant}.  Nevertheless, there are
many facets of theory, and various points of principle (some rather crucial), that 
have been insufficiently emphasized and articulated in detailed research papers, which of necessity 
present a given, though sometimes narrow, result. The upshot has been some defocusing of the theory 
enterprise and the accumulation of various myths that, while understood to be such by most practitioners, 
have at times confused the uninitiated.  With this paper, I highlight an eclectic mix of
important topics in core-collapse supernova theory that I feel have
not gotten sufficient ``air time." In the process, some of the central features
of modern supernova theory are identified.  It is hoped that this collection of excursions, 
though idiosyncratic, will help sharpen collective understanding of the core issues 
to be tackled on the way to a complete and credible understanding of the explosion phenomenon.

\section{Physical Context of Core-Collapse $-$ Basic Scenario}
\label{summary}

\subsection{Progenitors}
\label{progenitors}

A central facet of a star's evolution is the steady decrease with time
of its core specific entropy. The loss by outward diffusion of the energy 
generated by nuclear transmutation to progressively heavier elements
(for which nucleons are more and more organized in the nucleus) naturally leads 
to these lower entropies. This may seem counter-intuitive, since core temperature
increases during and between burning phases. However, core density also 
increases, and this increase outpaces the temperature increase needed to 
maintain burning and, as ash becomes fuel, to ignite the next burning stages. 
In addition, after the ignition of core carbon burning, the temperatures are 
sufficient to generate high fluxes of thermal neutrinos. These stream directly 
out of the core, accelerating core evolution and, in the sense of entropy, its 
refrigeration.  This phase also marks the evolutionary decoupling of the 
stellar photosphere from the burning core, which then inaugurates its more rapid
race to collapse. 

Entropy is relevant because it is a measure of randomness. A low value bespeaks 
organization and electron degeneracy. Hence, stellar evolution leads to white 
dwarf cores. The more massive stars generate more massive cores, and those 
more massive than $\sim$8 M$_{\odot}$ achieve the Chandrasekhar mass, at which
point such cores are unstable to the dynamical implosion (collapse) that 
inaugurates the supernova. However, the ``Chandrasekhar mass," and its core 
density profile when achieved, depend upon the electron fraction (Y$_e$) 
and entropy profiles.  The effective Chandrasekhar mass is not its canonical 
zero-entropy, uniform-Y$_e$ value of 1.456$(2{\rm Y}_e)^2$ M$_\odot$.  Entropy and Y$_e$ profiles are 
functions of the specific evolutionary paths to instability, in particular the character 
of convective shell burning and the $^{12}$C($\alpha$,$\gamma$)$^{16}$O rate. More 
massive progenitors evolve more quickly, and, therefore, don't deentropize as much by 
neutrino losses before collapse. They have higher entropies, resulting in additional 
pressure beyond that associated with zero-temperature electron degeneracy to support more mass, 
Therefore, the core mass necessary to go unstable is increased. Note that electron capture 
on nuclei alters Y$_e$, and the rates of electron capture depend sensitively upon isotope and density.

\begin{figure}
\centerline{
\includegraphics[height=0.80\textheight,angle=0]{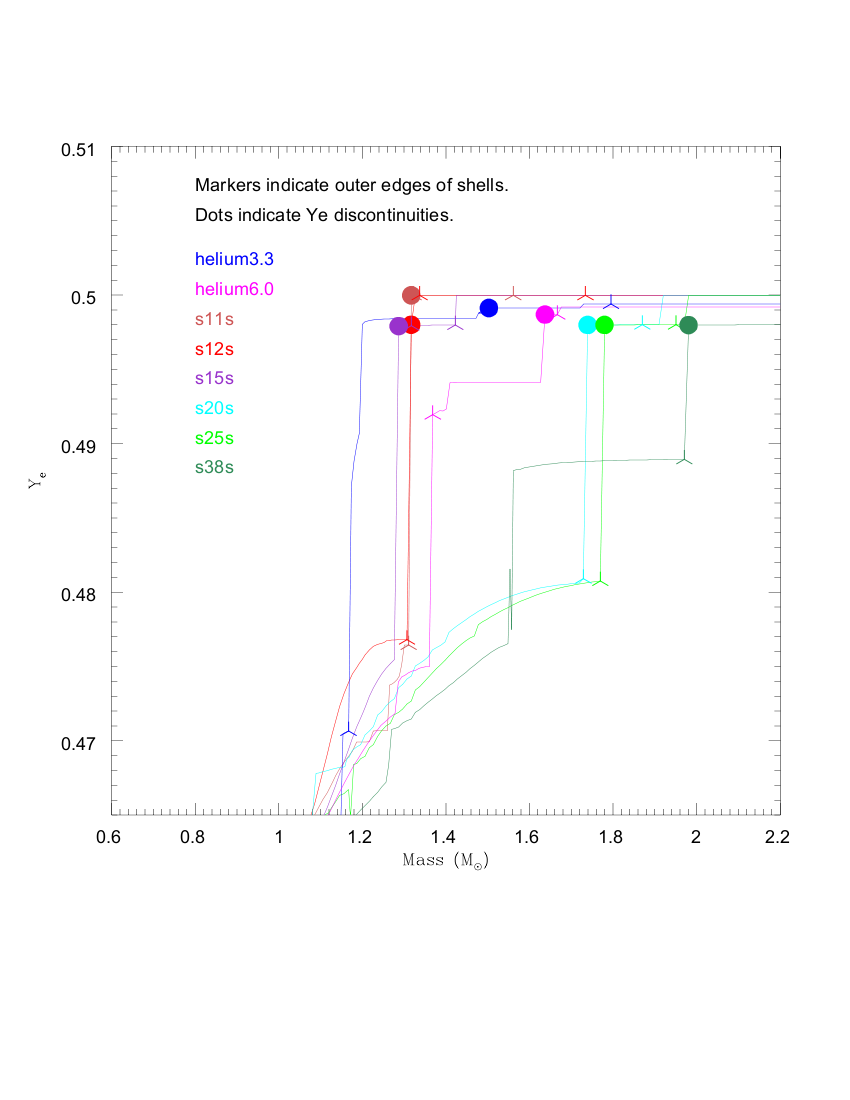}
}
\caption{The electron fraction, Y$_e$, versus enclosed mass for a suite of massive-star progenitor models 
from Woosley \& Weaver (1995) (``sXXs") and Nomoto \& Hashimoto (1988 $-$ ``heliumX" models).
The numbers given (the ``X"s in this caption) denote either the ZAMS masses or the helium-core masses.   
The dots indicate Y$_e$ discontinuities and the three-pointed markers indicate the boundaries of 
fossil burning shells. The outer edges of the iron cores are well-marked by the large circular colored dots.  
Note that they are positioned from $\sim$1.25 M$_{\odot}$ to $\sim$2.0 M$_{\odot}$ and are 
roughly in order of progenitor mass. See text for discussion.
\label{prog_ye}
}
\end{figure}

\begin{figure}
\centerline{
 \includegraphics[height=0.80\textheight,angle=0]{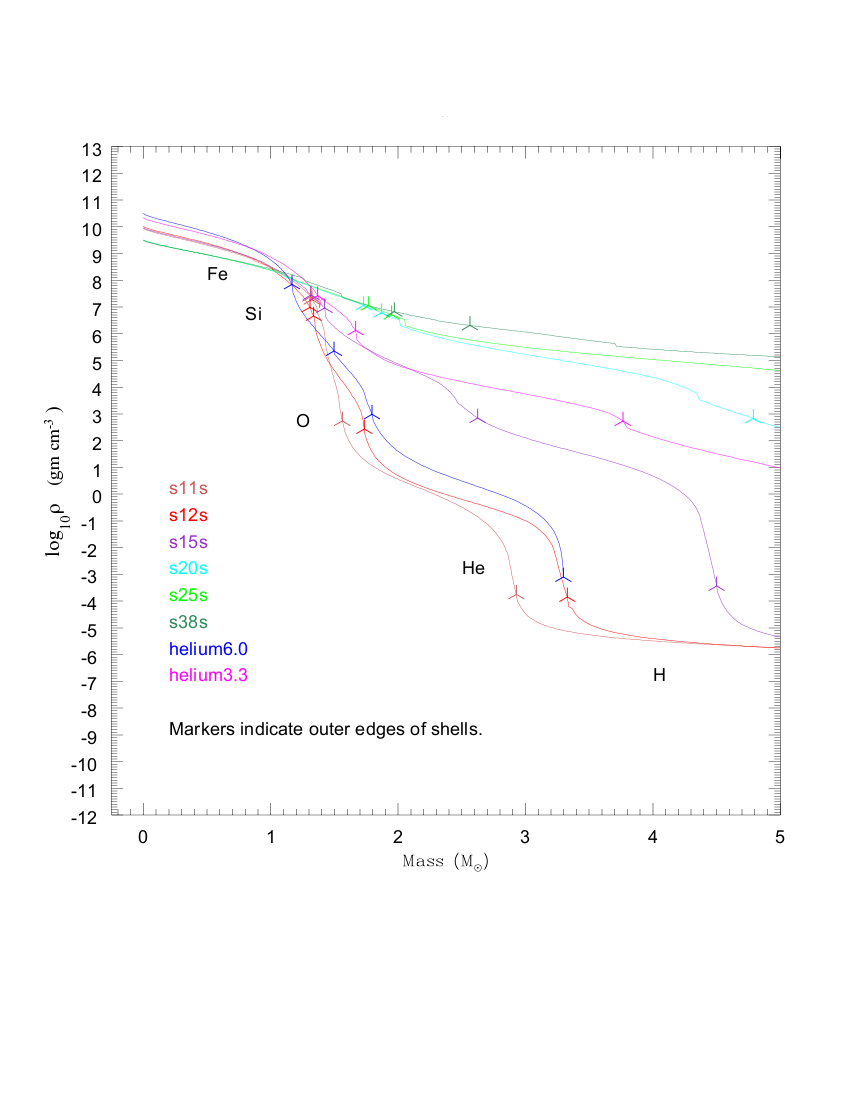}
}
\caption{Similar to Figure \ref{prog_ye}, but depicting the logarithm of the mass 
density ($\rho$, in gm cm$^{-3}$) versus interior mass (in solar masses) for 
various initial progenitor masses.  Note that the lower-mass progenitors have steeper slopes,
and that these profiles translate into more quickly dropping mass accretion rates ($\bar{M}$) 
through the stalled shock after bounce.
\label{prog_rho}
}
\end{figure}

The result is a ``Chandrasekhar mass" that could vary from $\sim$1.25 to 
$\sim$2.0 M$_{\odot}$, depending upon progenitor mass and evolutionary details (Figure \ref{prog_ye}).
The latter are not necessarily perfectly captured by current models. However, the
trend seems roughly to be that more massive progenitors have 1) larger 
effective Chandrasekhar masses and 2) envelopes in which the mass density 
decreases more slowly and, hence, that position more mass around the core.  Figure \ref{prog_rho} 
shows the core density profiles for various theoretical initial progenitor masses.
While most massive-star supernova progenitor cores evolve to iron peak elements and a Chandrasekhar-mass ``iron core" 
at the center of the canonical ``onion-skin" structure of progressively lighter 
elements from the inside out, the least massive progenitors (perhaps with ZAMS\footnote{Zero-Age-Main-Sequence} masses of 
$\sim$8.0$-$9.0 M$_{\odot}$) are thought to end up as ``O-Ne-Mg" cores (Nomoto \& 
Hashimoto 1988). Such cores might have very tenuous outer envelopes which, given 
current thinking, might result in underenergetic ($\sim$10$^{50}$ ergs $\equiv$ 
0.1 Bethe) neutrino-wind-driven explosions (Burrows 1987; Kitaura et al., 2006; Burrows et al. 2007c).
In both circumstances, the ashes from shell burning are responsible for fattening 
the inner core to the effective Chandrasekhar mass.  Importantly, the more massive 
progenitors have slightly lower central densities and temperatures at collapse. 
The higher densities of ``O-Ne-Mg" cores result in higher electron capture rates,
but the higher entropies of the more massive progenitors result in a greater 
softening of the EOS due to photodissociation\footnote{which decreases the 
effective $\gamma$ of the gas below the critical value of $4/3$}. Both processes 
facilitate the achievement of the Chandrasekhar instability, which once achieved is 
dynamical\footnote{The collapse time to nuclear densities is then 
no more than $\sim$350 milliseconds, whatever the progenitor mass. An approximate 
characteristic dynamical time is very roughly $\sim$${40 \ {\rm ms}}/{\sqrt{\rho_{10}}}$, where 
$\rho_{10}$ is in units of $10^{10}$ gm cm$^{-3}$.}. Therefore, 
in a real sense how the Chandrasekhar instability is achieved is 
secondary, and collapse proceeds similarly in both cases. Some would distinguish
``electron-capture supernovae" as a different species of supernova.  However, this
is really not justified.  The true difference is in the envelope density profile (Figure \ref{prog_rho}).
The lowest-mass massive progenitors have very steep density profiles outside the inner core.
These translate into sharply dropping accretion rates onto the proto-neutron star 
after bounce and before explosion. The dynamical differences and outcomes of ``iron-core"
or ``O-Ne-Mg" collapse are more dependent, therefore, on the different outer core mass
and density profiles, where the differences are expected to translate into real differences
in explosion energy, kick speeds, residual neutron star masses, and optical displays.
Since neither core type contains much thermonuclear fuel, 
unlike in the Type Ia supernova case of critical carbon-oxygen white dwarfs, burning does not 
inhibit collapse to nuclear densities ($\sim$$2.6\times 10^{14}$ gm cm$^{-3}$). However, 
the reader should note that, due to carbon and neon shell flashes, 
stellar evolution simulations up to the edge of collapse are very difficult 
for the lowest-mass massive stars and only one group has provided a massive-star 
model below 9.0 M$_{\odot}$ (Nomoto \& Hashimoto 1988). 

I reiterate that the initial core profiles, along with whatever initial rotation may 
be present in the core at collapse, must determine the spread in outcomes of collapse.  
In a very real sense, ``progenitor is destiny," a mapping complicated only by the 
randomness associated with chaotic turbulence and instability dynamics and with the 
unknown initial perturbation spectrum imposed by pre-collapse convective burning (Meakin \& Arnett 2006,2007ab). 
Due to such stochasticity, it is expected that a given progenitor star and structure will give 
rise to a distribution of outcomes (energies, proto-neutron star masses, kick speeds 
and directions, etc.), with a ``$\sigma$" that remains very much to be determined but 
that many hope (without yet much justification) will be small.  The task before 
theorists is to determine the progenitor/supernova mapping.  On the observational side, 
Smartt (2009) has recently attempted such a mapping by identifying progenitors 
in archival data to a handful of core-collapse supernovae.  His preliminary finding that no 
progenitor to a supernova with such archival data is more massive than $\sim$16 M$_{\odot}$
is intriguing, but will need further investigation to confirm or refute.      

With this background, core collapse proceeds (in theory\!) similarly for all progenitors. 
The inner $\sim$5000$-$10,000 kilometers is the most relevant.  Its dynamical time
is less than a second, while that of the rest of the star (with a radius of $\sim$10$^7 - 10^9$ km),
most of which is comprised of hydrogen and helium, is hours to a day.  Therefore, 
the dynamical inner core is decoupled from the outer shells and it is only 
when the supernova shock wave reaches them that they too participate
in the explosive dynamics, but then only as shocked spectators.  The initial core 
densities and temperatures are $\sim$6$-$15$\times 10^9$ gm cm$^{-3}$ and $\sim$6$-$10$\times 10^9$ Kelvin,
respectively. The initial core entropies are $\sim$0.7-1.2 $k_B$/baryon, where $k_B$ is Boltzmann's constant,
while the entropies in the outer fossil convective silicon and oxygen burning shells jump to $\sim$2$-$5.
There is a corresponding abrupt decrease in density at these shell boundaries, as well as increases in Y$_e$,
whose initial core values are $\sim$0.4$-$0.43. The outer shells are at lower densities, for which 
the electron capture rates are low and Y$_e$ is very near 0.5 (Figure \ref{prog_ye}).  The pressures are dominated 
by degenerate relativistic electrons, with slight thermal and Coulomb corrections, and 
for most progenitors most of the baryons are in nuclei near the iron peak. 
If we are dealing with an ``O-Ne-Mg" core, burning on infall will rapidly convert it into 
an ``iron-peak" core.  For an iron core, the nuclei are in nuclear statistical equilibrium,
which is a Saha equilibrium predominantly between nucleons in and out of nuclei, alpha particles, 
and the nuclei. Such a ``chemical" equilibrium is described, for a given nuclear model,
by temperature ($T$), mass density ($\rho$), and Y$_e$. 

\subsection{Collapse}
\label{collapse}

The instability that is collapse occurs because the average adiabatic $\gamma$ in the core 
is at or below $4/3$.  Photodissociation and domination by relativistic electrons guarantee this.
As collapse proceeds, $T$ and $\rho$ both rise. With the temperature increase, more nucleons
evaporate from nuclei.  If the number of free nucleons were to rise significantly, since they are 
non-relativistic ideal gases with a $\gamma$ of $5/3$, collapse would be halted and reversed 
before nuclear densities were achieved.  This was the supposition in the early 1970s.
However, as Bethe et al. (1979) and others have shown, the increase in temperature of 
the nuclei populates excited nuclear states, which represent many degrees of freedom. 
The upshot is a significant increase in the specific heat and the regulation of the temperature 
increase, since energy that would otherwise be channeled into kinetic degrees of freedom is redirected in part 
into these excited states. The result is not only moderation in the temperature increase during collapse,
but also in the production of non-relativistic free nucleons, thereby maintaining the domination of the pressure 
by the relativistic electrons and preserving the nuclei during collapse.  The consequence
is collapse all the way to nuclear densities, at which point nuclei phase transform 
into nucleons that at such densities experience strong nuclear repulsion, 
severely inhibiting further compression.  

\subsection{Bounce}
\label{bounce}

\begin{figure}
\centerline{
\includegraphics[height=0.70\textheight,angle=0]{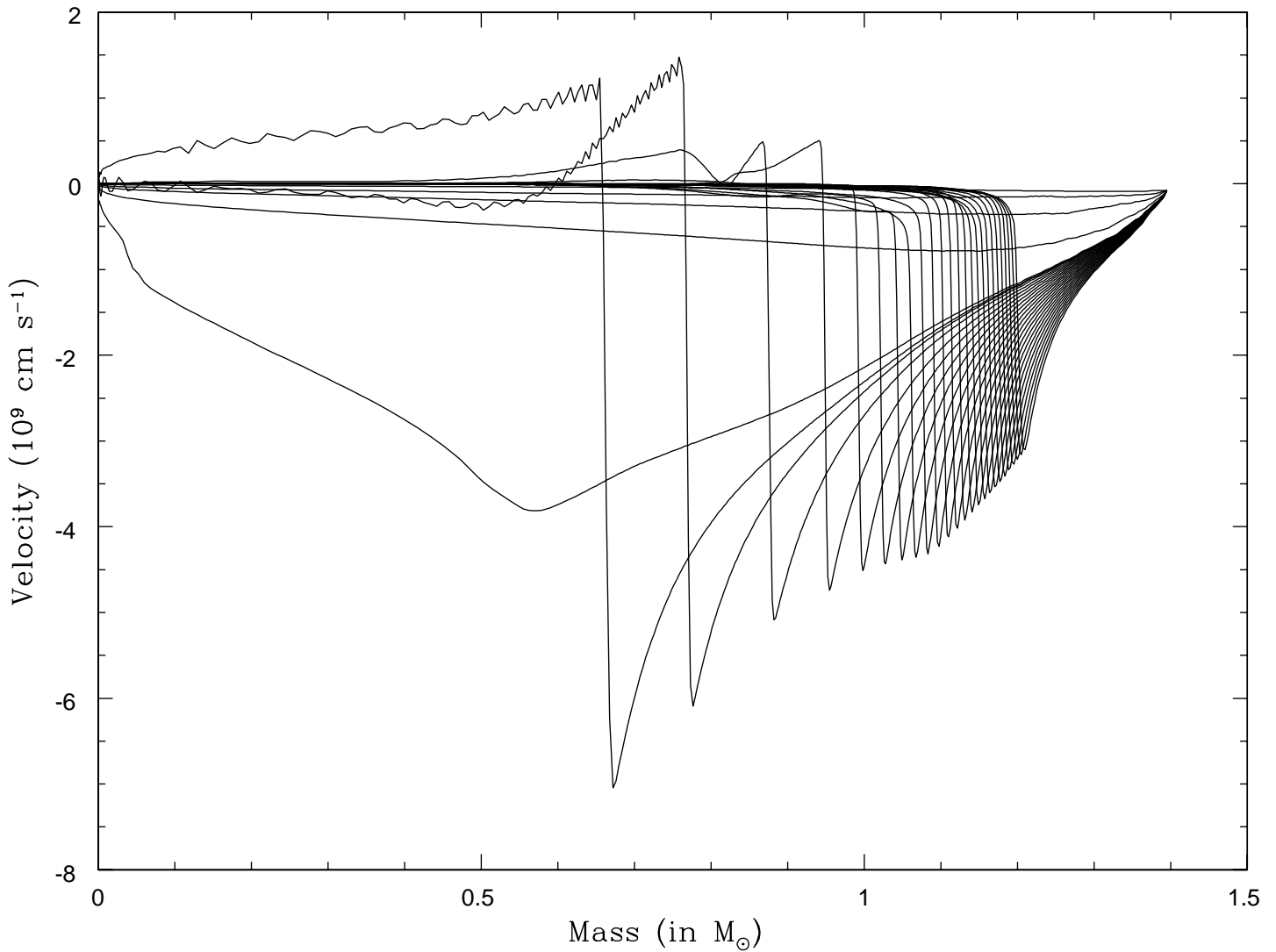}
}
\caption{The velocity (in $10^9$ cm s$^{-1}$) versus interior mass (in solar masses)
at various pre- and post-bounce times.  The shock wave, when present, is clearly indicated 
by the vertical drop, and is seen only for the post-bounce times. Within $\sim$20 milliseconds of bounce, the shock wave 
has stalled into accretion and post-shock speeds go negative. Reviving this structure is the 
goal of modern core-collapse supernova theory.  Note the line without the shock that changes 
slope in the middle.  This transition in slope near $\sim$0.6 M$_{\odot}$ marks the edge of 
the homologous core a few milliseconds before bounce.  Exterior to this minimum is the 
supersonic mantle whose maximum infall speed can reach $\sim$80,000 km s$^{-1}$. The 11-M$_{\odot}$ 
progenitor model from Woosley \& Weaver (1995) was used for this Figure, as well as for Figures 
4, 6, 8, \& 9. (Taken from numerical data generated in Thompson, Burrows, \& Pinto 2003.)
\label{vel_11}
}
\end{figure}

Within less than a millisecond, this stiffening of the equation of state  
halts and reverses collapse.  During collapse, since the central speed and outer 
core speeds must be zero and low (respectively) the peak collapse speed is 
achieved in the middle of the collapsing core.  Early during collapse, this results 
in a separation into a subsonic inner core, which collapses almost homologously 
($v \propto r$) and as a unit, and a supersonically infalling outer core. The peak
speed of the inner shells of the outer core are roughly a constant large fraction of free-fall.  
Therefore, when the inner core achieves nuclear densities and rebounds, because these two regimes
are out of sonic contact, the subsonic inner core bounces as a unit and as a spherical piston
into the outer core, which is still collapsing inward, thereby generating a shock wave
at the interface (Figure \ref{vel_11}). This is the supernova shock in its infancy. 

During collapse, increasing density and temperature result in increasing electron capture rates
on both nuclei and free protons, with the resulting decrease in Y$_e$.  Current thinking is that 
capture on nuclei predominates (Langanke \& Martinez-Pinedo 2003), but since the ``iron peak" shifts 
with the increase in $\rho$ and the decrease in Y$_e$ to higher and higher atomic weights, 
exotic isotopes, for which we have no data, quickly dominate. However, this uncertainty 
alters the progress of collapse only slightly, since gravitational free fall bounds
collapse speeds, whatever the capture rates.  The result is that pre-bounce collapse is
universal in character, requiring (after peak speeds have achieved $\sim$1000 km s$^{-1}$)
between $\sim$150 and $\sim$350 milliseconds to achieve bounce.  More importantly, rewinding 
collapse from bounce yields almost universal trajectories relative to bounce time.    

\subsection{Trapping}
\label{trapping}

As indicated, electron capture during collapse lowers Y$_e$, but it also produces electron-type 
neutrinos ($\nu_e$) at progressively increasing rates.  The average energy of these $\nu_e$s
increases with density and temperature.  Since the cross section for scattering off nuclei
by the coherent process (Freedman 1974) increases roughly as the square of 
neutrino energy and the densities are soaring at a rapid rate, the mean-free-paths for $\nu_e$-matter
interactions are fast decreasing.  When these mean-free-paths become comparable to the scale-height
of the matter density or when the outward diffusion speed of the $\nu_e$s equals the infall speed, 
then the $\nu_e$s are {\it trapped} in the flow (Mazurek 1974; Sato 1975). After trapping, 
electron capture is balanced by $\nu_e$ capture to establish chemical equilibrium at a given 
electron lepton fraction.  

Since this condition is achieved not long after the central density 
is $\sim$10$^{11}$ gm cm$^{-3}$, trapping of electron lepton number and $\nu_e$s happens before 
much electron capture has occurred and has a profound effect on collapse and subsequent evolution. Trapping 
locks electron lepton number, electrons, and $\nu_e$s in the core for many seconds, depending 
on the mass shell. The trapped $\nu_e$s are compressed significantly, but at low entropy 
and conserving lepton fraction. The latter settles near 0.30 (partitioned between 
electrons [$\sim$0.25] and $\nu_e$s [$\sim$0.05]), instead of 0.03 (all electrons), which 
it might have been without trapping.  The result is the channeling by compression 
of a significant fraction of the gravitational energy of collapse into {\it degenerate} $\nu_e$s and electrons,
with central chemical potentials and Fermi energies of $\sim$100$-$300 MeV.  Hence, trapping is not merely
the increase of the $\nu_e$ optical depth far beyond values of one, but the preservation at high values (not
far from the initial value of $\sim$0.43) of the lepton fraction and electron numbers in the core
and the production of a sea of degenerate $\nu_e$ whose average energy, by dint of compression 
after trapping, is high. If it weren't for lepton trapping, the $\nu_e$s would be thermal with
average energies at bounce of $\sim$30 MeV, optical depths of hundreds, and energy diffusion times of 
$\sim$50-100 milliseconds.  Instead, their inner core average energies are 100-300 MeV, their optical
depth to infinity from the center is $\ge$10$^5$, and the energy diffusion time out of the core 
is many seconds.  The latter has been boosted by the compressional increase in the average $\nu_e$ energy,
for which the $\nu_e$-matter cross sections are much larger than they would otherwise be. 

By keeping Y$_e$ high, trapping thwarts the rapid evaporation of neutrons from the nuclei 
that would otherwise be too neutron-rich to retain them. Therefore, both the increased specific heat 
due to excited nuclear states and trapping ensure the survival of the nuclei and the 
maintenance of the effective $\gamma$ below $4/3$ until nuclear densities are achieved. 
If it weren't for trapping and excited nuclear states, bounce would occur at sub-nuclear 
densities, the average core electron neutrino energy would be much lower, and $\nu_e$-matter mean-free-paths
at bounce would be much longer.  The lower bounce densities and longer mean-free-paths
would have translated directly into much shorter deleptonization and cooling times for a proto-neutron 
star. Hence, trapping is centrally important in explaining the long duration of the neutrino burst from
SN1987A (Kamioka II $-$ Hirata et al. 1987; IMB $-$ Bionta et al. 1987).  Before neutrino trapping was recognized,
the energy diffusion time out of the proto-neutron star was thought to be $\sim$100 milliseconds,
a factor of $\sim$100 shorter than observed.  Moreover, since the total reservoir of energy
radiated is fixed at the gravitational binding energy of a cold neutron star 
($\sim$$\frac{GM^2}{R} \sim 3\times 10^{53} {\rm ergs} \equiv$ 300 Bethes), 
a short duration would have implied a higher average neutrino energy ($\ge$50 MeV)
than measured by Kamioka ($\sim$15 MeV $-$ for the $\bar{\nu}_e$s). Therefore, 
and importantly, the lower measured energy and higher measured duration are direct 
consequences of electron neutrino (lepton number) trapping, a fact not widely 
appreciated.

Trapping halts the short-term decrease in Y$_e$ during collapse, but, as mentioned, there 
is still a slight decrease to $\sim$0.25$-$0.35.  The magnitude of this decrease is enhanced
by inelastic scattering of $\nu_e$s off electrons (Mezzacappa \& Bruenn 1993abc). Like Compton 
scattering, the capture-produced $\nu_e$s are downscattered in energy and at lower energies 
their Freedman scattering cross sections are lower. The result is a slightly slower 
increase in the optical depth during infall, and this results in slightly delayed trapping 
at lower Y$_e$s. The magnitude of the effect is $\sim$0.02-0.03.   

\begin{figure}
\centerline{
\includegraphics[height=0.70\textheight,angle=0]{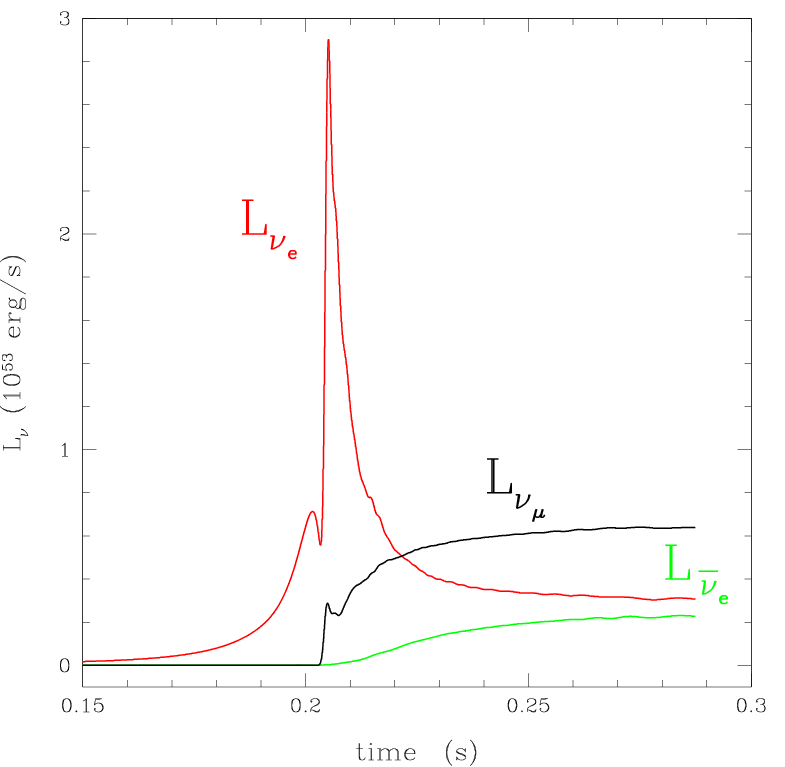}
}
\caption{The luminosity (in units of $10^{53}$ ergs s$^{-1}$) at infinity of the 
$\nu_e$, $\bar{\nu_e}$, and, collectively, the $\nu_{\mu}$, $\bar{\nu}_{\mu}$, $\nu_{\tau}$, 
and $\bar{\nu}_{\tau}$ neutrinos, versus time (in seconds) around bounce.  The $\nu_e$
breakout burst is clearly seen.  After shock breakout, the temperatures are sufficient
to generate the other species in quantity.  Generally, the non-electron types
carry away $\sim$50\% of the total, with the $\nu_e$s and $\bar{\nu_e}$s sharing the rest.
(Plot taken from Thompson, Burrows, \& Pinto 2003.)
\label{lum_time}
}
\end{figure}

The trapped Y$_e$ sets the mass scale of the homologously-collapsing inner core at approximately the 
associated Chandrasekhar mass (Yahil 1983; Burrows \& Lattimer 1983).  
Since M$_{Ch}$ is proportional to ${\rm Y_e}^2$, this is $\sim$0.5$-$0.6 M$_{\odot}$.  Therefore,
the bounce shock first emerges at the sonic point near $\sim$0.5$-$0.6 M$_{\odot}$ ($\sim$10-20 km)
between the inner core and the outer supersonic mantle in an optically thick region.
It then propagates down the density gradient, entropizing the matter it encounters, dissociating the nuclei
into nucleons, and reaching lower $\nu_e$ optical depths within $\sim$1 millisecond.  At this point,
the copious sea of $\nu_e$s, newly-liberated by electron capture via the super-allowed 
charged-current capture process on newly-liberated protons, ``breaks out" 
in a burst that lasts $\sim$10 milliseconds (Figure \ref{lum_time}).  The luminosity of 
this electron neutrino breakout burst is within an order-of-magnitude 
of the total optical output of the observable Universe ($\sim$$3-4 \times 10^{53}$ ergs s$^{-1}$), 
and is the first, most distinctive, feature of the supernova neutrino emission process.  It is 
a firm prediction of generic core-collapse supernova theory, and if it doesn't exist, then much of
the supernova theory developed in the last $\sim$35 years is wrong. 

\subsection{The Problem}
\label{problem}

\begin{figure}
\centerline{
 \includegraphics[height=0.90\textheight,angle=0]{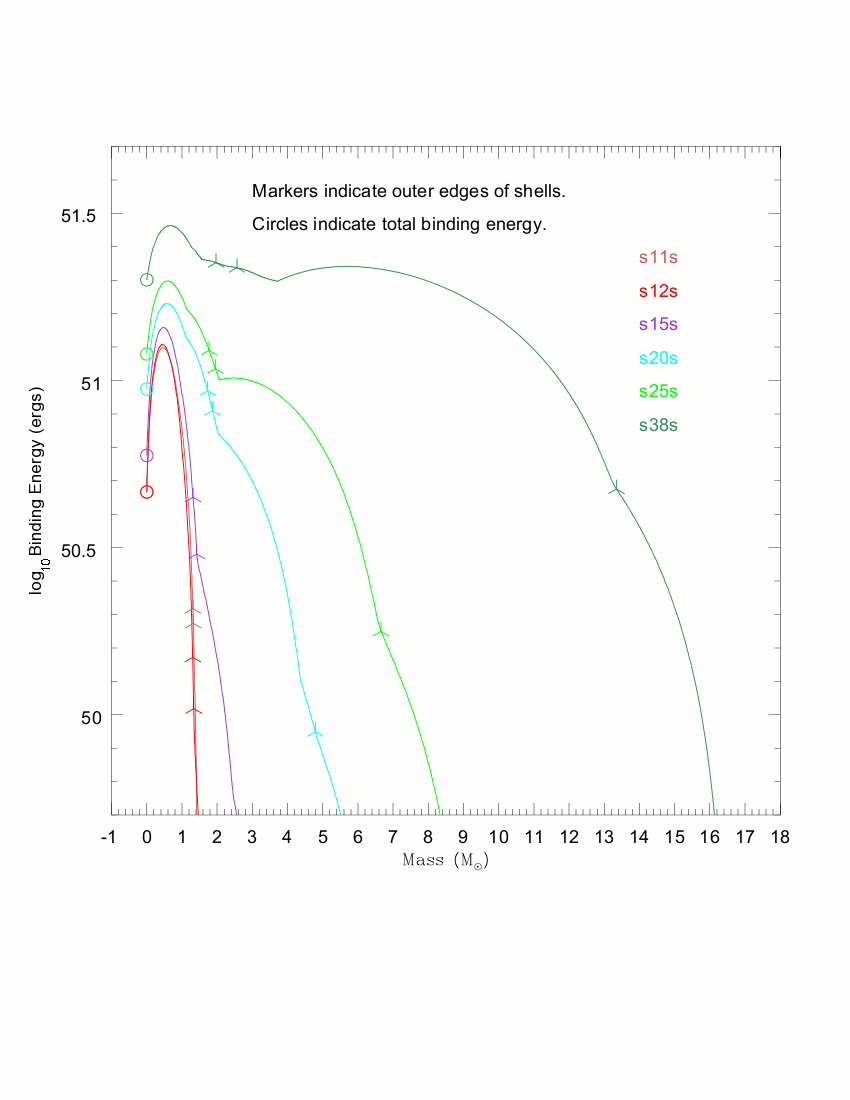}
}
\caption{The logarithm base ten of the gravitational binding energy (in ergs, including 
the thermal energy) of the shells in various progenitor stars (see Figure \ref{prog_ye}) 
exterior to the interior mass coordinate (in M$_{\odot}$) shown on the abscissa. These 
are the approximate energies that the supernova blast must overcome to eject the 
stellar shells exterior to a given residual neutron star or black hole mass. 
Note that the more massive the progenitor the greater the binding energy cost
for a given baryon mass left behind. 
\label{bind}
}
\end{figure}

The direct mechanism of explosion posits that this bounce shock continues unabated 
outward into the star and is the supernova.  However, both simple theory and detailed 
numerical simulations universally indicate that the $\nu_e$ burst and photodissociation 
of the infalling nuclei debilitate the shock wave into accretion within $\sim$5 milliseconds 
of bounce. In a very real (though approximate) sense, the subsonic inner core and shocked mantle 
together execute a coherent harmonic oscillation that is near-critically damped. 
The shock acts like a black-body absorber of sound (the bounce pressure pulse), 
and the breakout neutrinos and photodissocciation do the rest.  The 
breakout neutrino burst directly saps the shock of energy, while photodissociation
by the shock redistributes shock energy from the kinetic component (and, hence, pressure)
to pay the nuclear binding energy penalty.  Contrary to common lore, photodissociation
is not a loss of energy $-$ the energy is still there and could be recovered with recombination.
Rather, photodissociation softens the equation of state by lowering the $\gamma$ and raising
the specific heat. The result is a less efficient conversion of infall kinetic energy (whose original
source is gravity) into pressure.  One should note that if electron capture and neutrino transport
are both artifically turned off during and after collapse, but a realistic EOS with photodissociation
is employed, the direct mechanism can be shown to work for many of the published progenitor 
models.  However, even then, with Y$_e$ frozen at its initial value and no neutrino burst or losses,
the energy of the explosion is never higher than a few tenths of a Bethe, not accounting for 
the need to overcome the gravitational binding energy of the rest of the star above 
a ``canonical" neutron star mass cut of $\sim$1.5 M$_{\odot}$ (Figure \ref{bind}).  Something more is needed.

Therefore, since circa 1980 theorists have been presented with a stalled accretion 
shock at a radius near $\sim$100-200 km and have been trying to revive it.  This was 
and is an unsatifactory state of affairs. Supernova rates, O- and B-star death rates, and
neutron star birth rates all suggest that most massive stars explode as supernovae and
leave neutron stars. The fraction that branch into the stellar-mass black hole
channel is unknown. The fraction of times stellar-mass black hole formation is accompanied
by a supernova is unknown.  If the shock is not revived and continues to accrete, all 
cores will collapse to black holes.  The serious revival mechanisms include the delayed 
neutrino mechanism (Wilson 1985), magnetohydrodynamic (MHD) bipolar jet production 
(LeBlanc \& Wilson 1970; Bisnovatyi-Kogan et al. 1976; Symbalisty 1984; Burrows et al. 2007d; Takiwaki \& Kotake 2011) (requiring very 
rapid rotation), and the acoustic mechanism (Burrows et al. 2006,2007b).  The spherical delayed neutrino 
mechanism works only weakly, and then only as a wind (Burrows 1987; Kitaura et al. 2006; 
Burrows et al. 2007c) for one non-representative progenitor (Nomoto \& Hashimoto 1988), 
but the multi-dimensional variant is considered the frontrunner in the generic 
progenitor case. However, using the most advanced multi-D numerical codes and incorporating
detailed microphysics, no one has yet been able to explain or reproduce any of the 
five ``facts" of core-collapse supernova listed at the beginning of the 
introduction\footnote{except, perhaps, pulsar kicks: Scheck et al. 2004, 2006; 
Wongwathanarat et al. 2010; Nordhaus et al. 2011}. 

\begin{figure}
\centerline{
\includegraphics[height=0.90\textheight,angle=0]{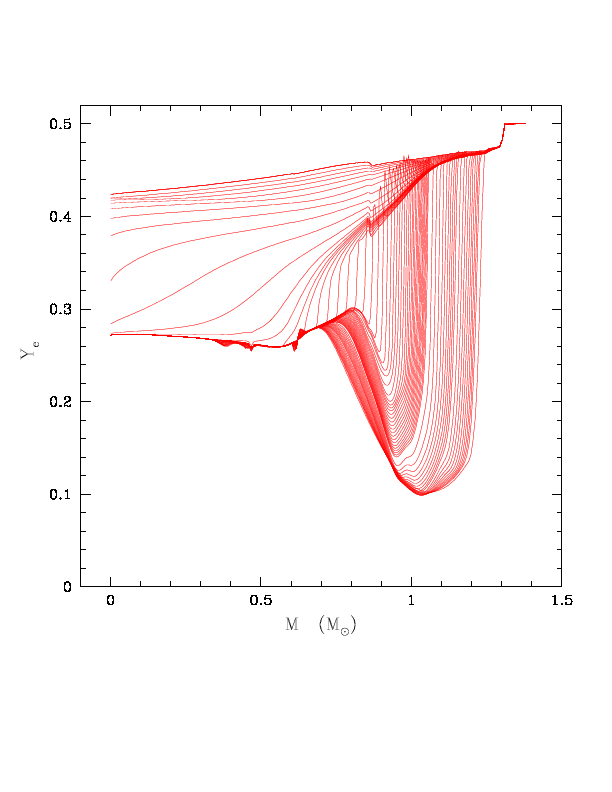}
}
\caption{Snapshots of Y$_e$ versus interior mass (in solar masses) profiles before
and just after bounce.  Capture decreases Y$_e$  on infall, but lepton number is soon trapped
in the interior.  After bounce, the outward progress of the shock wave liberates
$\nu_e$ neutrinos, creating a trough in Y$_e$ interior to the shock, but exterior
to the opaque core.  The shock wave resides just rightward of the steep drop in Y$_e$ 
on the right. The numerical data for this 11 M$_{\odot}$ progenitor model run were 
taken from Thompson, Burrows, \& Pinto (2003).
\label{ye_11}
}
\end{figure}

The stalled shock continues to dissociate infalling nuclei into nucleons. 
Electron capture onto the newly-produced nucleons at densities of $\sim$10$^9$ $-$ 10$^{10}$ gm cm$^{-3}$ and 
with temperatures of $\sim$1.5 $-$ 2.5 MeV rapidly lowers the post-shock Y$_e$ to values
of $\sim$0.1-0.2, creating the characteristic Y$_e$ trough seen in all detailed simulations 
(Figure \ref{ye_11}). Due both to electron capture behind the shock and neutrino cooling 
from the neutrinospheres\footnote{at average radii of $\sim$20-60 km, densities of $\sim$10$^{11}$ 
to $\sim$10$^{12}$ gm cm$^{-3}$, and temperatures of $\sim$4$-$7 MeV of all six neutrino types 
($\nu_e$, $\bar{\nu}_e$, $\nu_{\mu}$, $\bar{\nu}_{\mu}$, $\nu_{\tau}$, 
and $\bar{\nu}_{\tau}$)} neutrino and electron-lepton-number losses continue apace.
The proto-neutron star becomes more and more bound. 

\begin{figure}
\centerline{
\includegraphics[height=.80\textheight,angle=0]{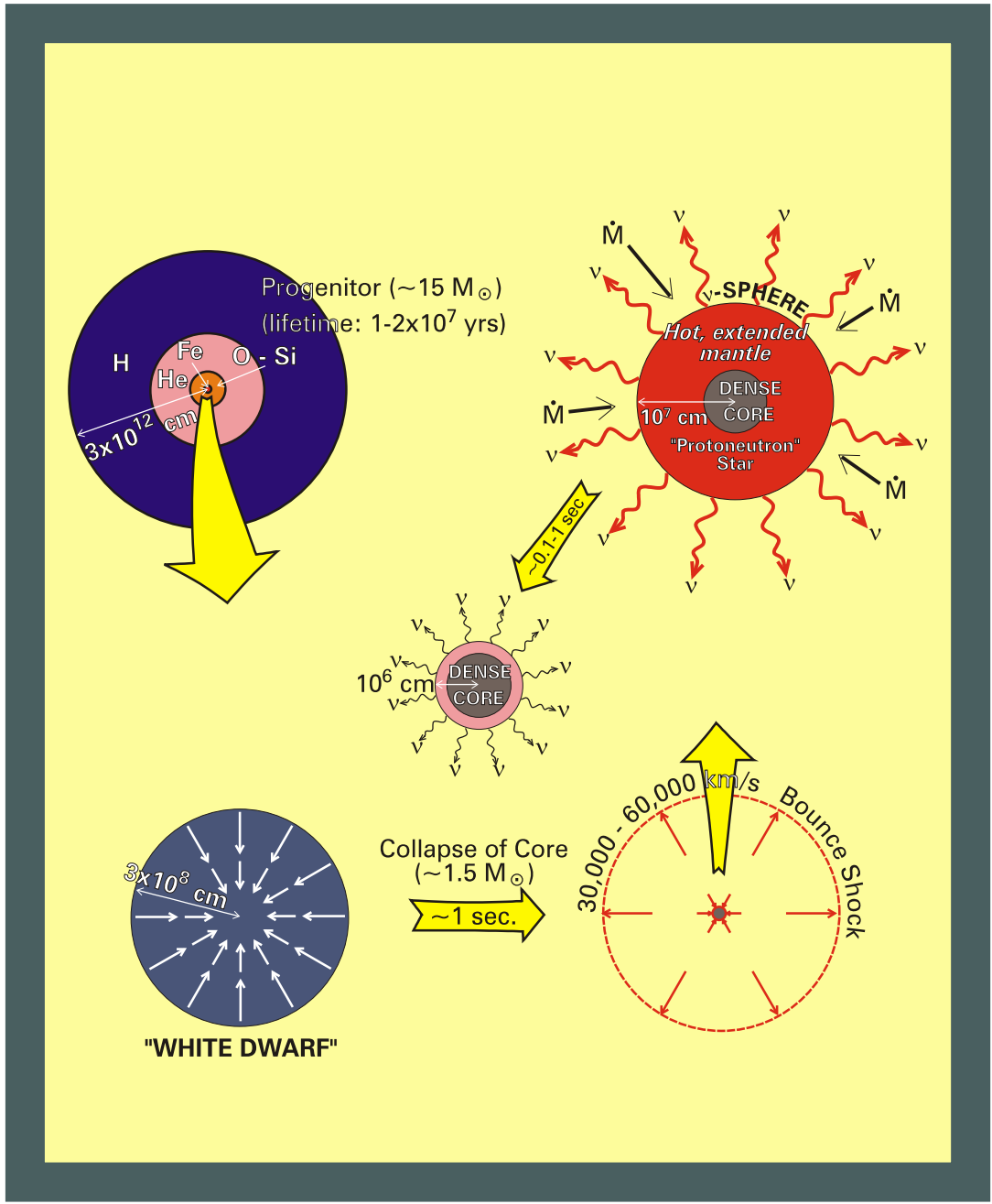}
}
\caption{A storyboard of the evolution of the core of the ``onion-skin" structure 
into the radiating proto-neutron star.  See text for a discussion.
\label{circles}
}
\end{figure}

\begin{figure}
\centerline{
\includegraphics[height=0.70\textheight,angle=0]{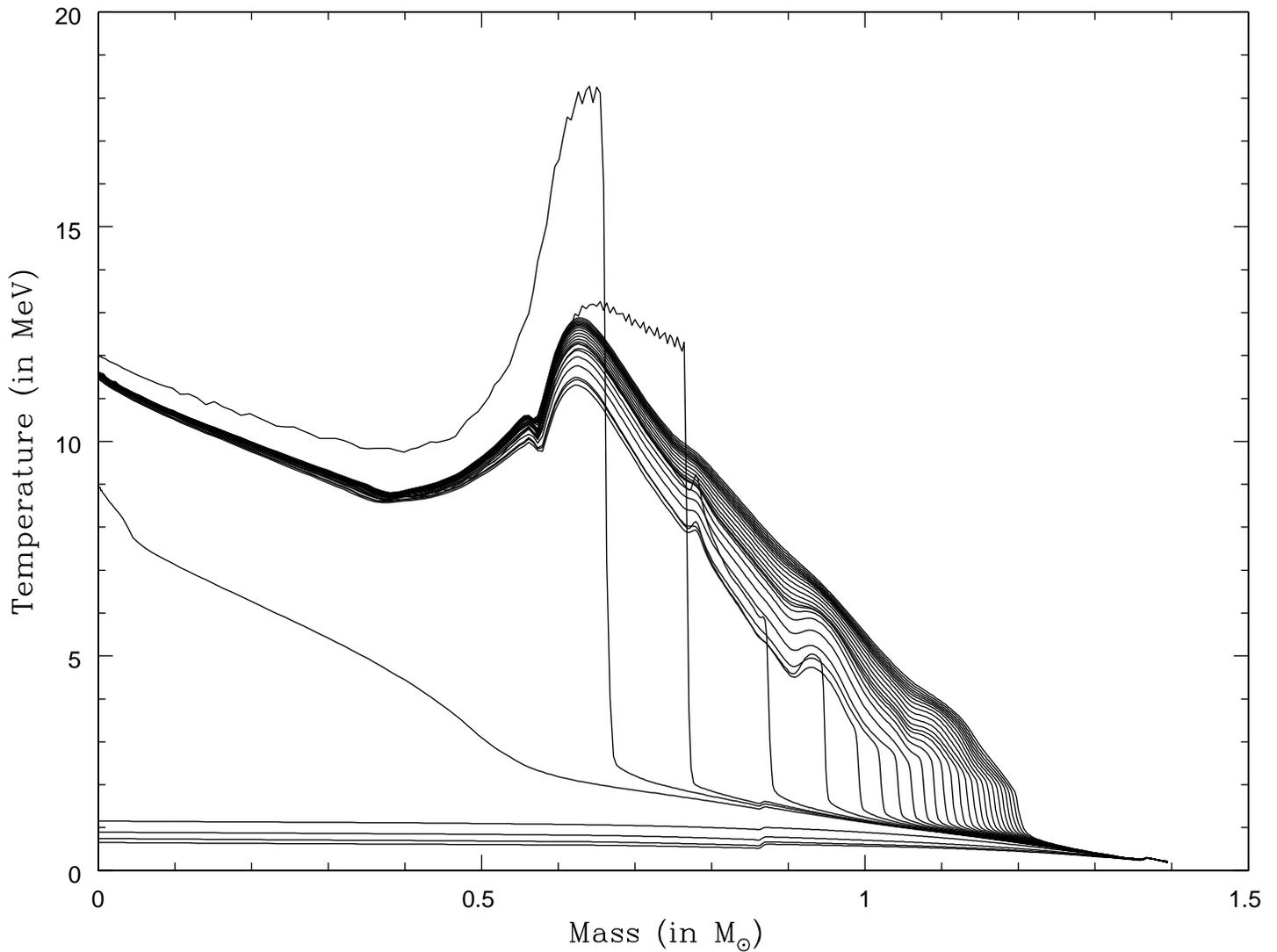}
}
\caption{Temperature (in MeV) versus interior mass (in solar masses) at various times
before and just after bounce.  The shock is depicted by the vertical drop in temperature 
and propagates out in mass as (and after) it stalls into accretion.  Initial post-bounce 
central temperatures are $\sim$10 MeV and the peak temperature just before shock 
breakout can exceed 20 MeV, but soon falls. The various neutrinopheres after 
$\sim$20$-$100 milliseconds after bounce are at $\sim$1.0$-$1.1 M$_{\odot}$. 
(Numerical data taken from Thompson, Burrows, \& Pinto (2003).)
\label{temp_11}
}
\end{figure}

\begin{figure}
\centerline{
\includegraphics[height=0.70\textheight,angle=0]{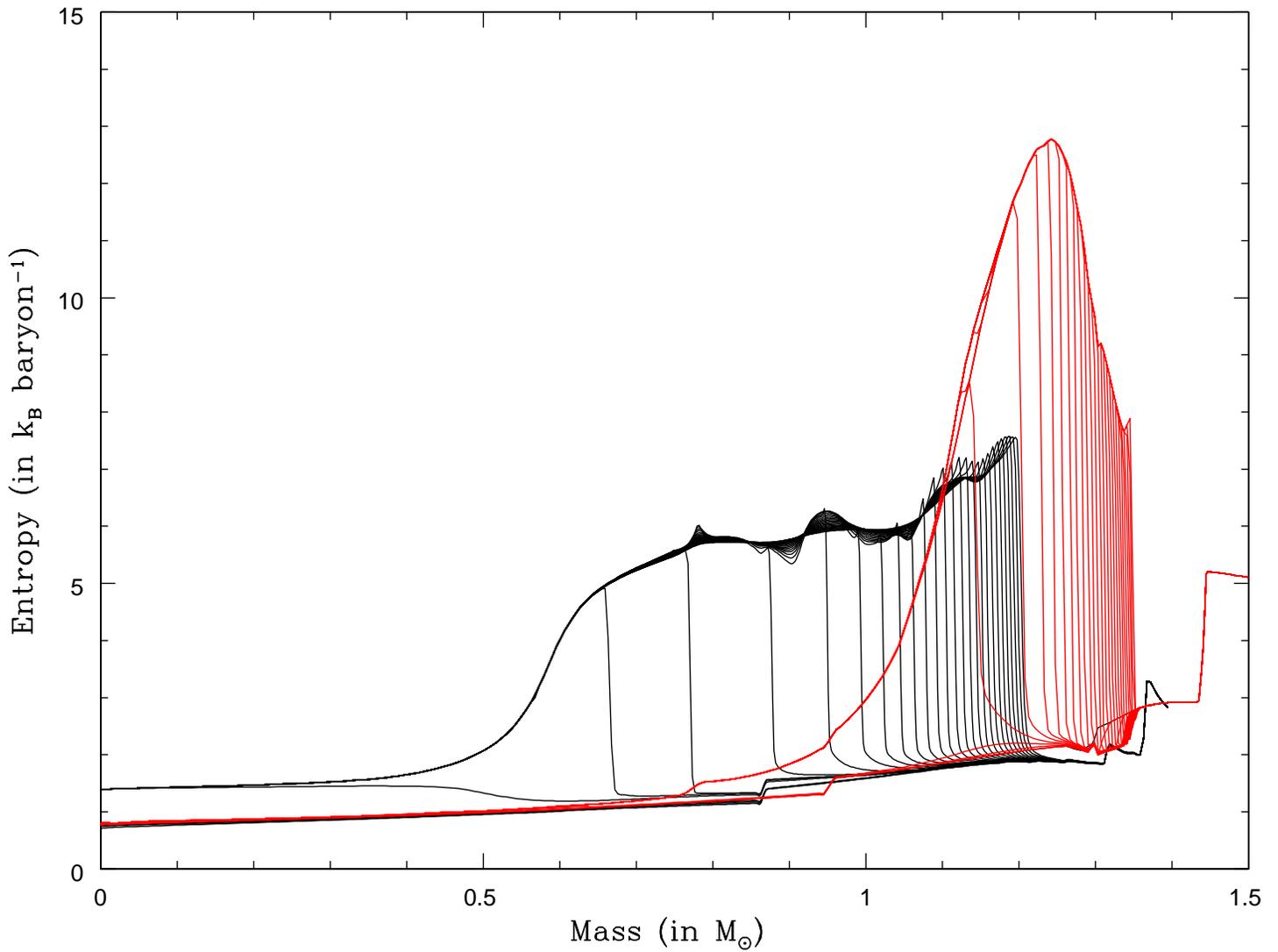}
}
\caption{Similar to Figure \ref{temp_11}, but for entropy (in $k_B$ per baryon) versus interior mass
(in solar masses). The black curves depict the evolution of profiles using realistic neutrino 
transport and stopping $\sim$10 milliseconds after bounce, while the red curves depict the 
corresponding developments when neutrino physics is turned off.  Note that the red 
curves extend further in mass and reach much higher early mantle entropies. See text for discussion.
(Numerical data taken from Thompson, Burrows, \& Pinto (2003).) 
\label{ent_11.d}
}
\end{figure}

Therefore, the object of current intense investigation is a quasi-static proto-neutron star 
(Burrows \& Lattimer 1986; Figure \ref{circles}) with a baryon mass of 
$\sim$1.2$-$1.5 M$_{\odot}$ (depending upon progenitor), an initial central density
at or above $\sim$$4.0 \times 10^{14}$ gm cm$^{-3}$ and rising, a central temperature of $\sim$10 MeV (Figure \ref{temp_11}) and rising, 
an inner core entropy of $\sim$1, and a shocked outer mantle entropy with a peak value ranging 
near $\sim$6$-$15 (Figure \ref{ent_11.d}), all bounded by an accretion shock wave stalled near 
$\sim$100$-$200 km. During the delay to explosion, the accretion rate ($\dot{M}$) decreases 
from $\sim$1.0 to 0.3 M$_{\odot}$ per second (for the most massive progenitors) to from 
$\sim$0.4 to 0.05 M$_{\odot}$ per second (for the least massive progenitors with 
iron cores). The associated accretion ram pressure, post-shock 
electron capture, and deep gravitational potential well are major impediments to explosion.
If there were no accretion, neutrino heating would explode the proto-neutron star mantle 
immediately. The rate of accretion and its evolution are functions strictly of the progenitor mass 
density profile just prior to collapse, which is a central determinant of the outcome.

\section{The Current Status of Core-Collapse Simulations}
\label{status}

There has been palpable progress in the development of techniques and tools
to address the core-collapse problem in the last thirty years, but the
current status of the theory for the mechanism and the systematics of
core-collapse explosions is ambiguous, if not confusing.  Wilson (1985),
in a pioneering paper and using a spherical code, obtained a neutrino-driven
explosion after a short post-bounce delay, but only if he included a mixing-length
algorithm to mimic doubly-diffusive ``neutron-finger" instabilities that dredged up
heat and thereby enhanced the driving neutrino luminosity and heating
in the ``gain region" (Bethe \& Wilson 1985).  The gain 
region is the region behind the shock in which there is net neutrino 
heating. Without this boost, he did not obtain explosions. 
However, Bruenn \& Dineva (1996) and Dessart et al. (2006) have 
shown that such instabilities don't arise in proto-neutron stars.
More recently, Kitaura et al. (2006) and Burrows et al. (2007c) obtained weak 1D (spherical)
explosions ($\sim$10$^{50}$ ergs) via a neutrino-driven wind for the lowest
mass progenitor in the literature ($\sim$8.8 M$_{\odot}$). These are the only
models credibly shown to explode in 1D (Mezzacappa et al. 2001). This
low explosion energy comports with the inference that the energy of
explosion may be an increasing function of progenitor ZAMS mass (Utrobin \& Chugai 2009).
Moreover, explosion occurred before convective overturn instability could be obvious.
This result's 1D simplicity lends credence to these state-of-the-art simulations using the best physics.
However, this progenitor has a uniquely steep density gradient
just exterior to the core (and, hence, almost no accretion tamp), and such progenitor
density profiles and 1D explosion behavior are seen in no other circumstance.
This highlights an important point. There is not just one
``core-collapse supernova problem,'' but many. The properties of
presupernova stars vary and so too will the explosion model. More
massive stars are more difficult to explode and at some point probably 
make black holes (Ugliano et al. 2012; O'Connor \& Ott 2011; Fryer et al. 1999). 

Shock-imposed negative entropy gradients at bounce, neutrino heating from below,
the standing-accretion-shock-instability (``SASI"; Blondin, DeMarino, 
\& Mezzacappa 2003;  Foglizzo, Scheck, \& Janka 2006), and negative lepton gradients 
in the inner core all render the shock-bounded environment of the proto-neutron star
unstable and turbulent, severely breaking spherical symmetry in the general case.
Indeed, it was shown some time ago that multi-dimensional instabilities obtain
and are probably central to the core-collapse supernova mechanism in most cases
(Burrows \& Fryxell 1992; Herant et al. 1992,1994; Burrows, Hayes \& Fryxell 1995; Janka \& M\"uller 1996;
Blondin, DeMarino, \& Mezzacappa 2003; Fryer \& Warren 2002,2004; Foglizzo, Scheck, 
\& Janka 2006). The salient processes in the neutrino heating model may be the 
enhancement of the neutrino-matter heating efficiency, the turbulent pressure, and the 
enlargement of the gain region, among others. Neutrino heating in the shocked mantle exterior 
to the neutrinospheres is probably the foremost driver of convection 
(Burrows, Dolence, \& Murphy 2012; Murphy et al. 2012).  
This is an unexceptional conclusion, since in the neutrino heating mechanism, it is the 
driver of explosion and the major source of explosion energy. 

Using detailed neutrino transport and 2D hydro, Marek \& Janka (2009)
observed an explosion of a 15-M$_{\odot}$ progenitor via the turbulent
neutrino heating mechanism. However, it seems underpowered ($\le$10$^{50}$ ergs)
for this ``average" progenitor, and explosion was not seen for their higher-resolution run.
Recently, M\"uller (2011) and M\"uller, Janka, \& Marek (2012) obtained explosions in 2D of 
both 11.2 M$_{\odot}$ and 15-M$_{\odot}$ progenitors using state-of-the-art neutrino 
transport with conformal general relativity (emphasizing the importance of the 
latter; see also Kuroda, Kotake, \& Takiwaki 2012), but the explosion energies were 
$\sim$2.5$\times$10$^{49}$ and $\sim$10$^{50}$ergs, respectively, at the end of the calculations. When 
account is taken of the binding energy of the overlying star (Figure \ref{bind}) these 
energies could be negative, potentially undermining these state-of-the-art 2D simulations 
as viable supernova explosion models. However, the asymptotic energies remain to be determined
and it should be acknowledged that these are the best 2D simulations of collapse yet performed.
However, as in Marek \& Janka (2009), though the hydro was 2D, the transport in the 
M\"uller (2011) and M\"uller, Janka, \& Marek (2012) papers was done multiple times in 1D along 
numerous radial rays.  The so-called ``ray-by-ray" formalism uses the temperature ($T$),
density ($\rho$), and electron fraction (Y$_e$) profiles along a given radial ray to feed a 1D spherical transport
calculation, that then provides local heating rates.  Since vigorous convection in 2D
leads to significant angular variation in $T$, $\rho$, and Y$_e$, this approach exaggerates the
variation in the neutrino heating with angle. Ott et al. (2008), performing the only multi-group, multi-angle
neutrino transport calculations ever done in 2D, have shown that the integral
nature of transport smoothes out the radiation field much more than the matter fields.
This suggests that local neutrino heating rates obtained using the ray-by-ray approach
might unphysically correlate with low-order mode shock and matter motions, and could be pumping them
unphysically. This may contribute to the perception that the $\ell = 1$ dipolar shock oscillation
seen in such 2D simulations is essential to the turbulent, multi-D neutrino-driven mechanism.

Burrows et al. (2006,2007b) performed 2D radiation/hydro multi-group, flux-limited
simulations and did not obtain neutrino-driven explosions.  Their algorithm was Newtonian 
and did not include velocity-dependent, general-relativistic, or inelastic scattering terms 
in the transport.  Nevertheless, $\sim$1 second after bounce they observed 
vigorous inner core g-mode oscillations that generated a sound pressure 
field sufficient to explode the envelope.  However, the energy of explosion
via this ``acoustic" mechanism was very slow to accumulate,
reaching only $\sim$10$^{50}$ ergs after more than $\sim$0.5 seconds.  In addition,
Weinberg \& Quataert (2008) suggest that the amplitude of these g-mode oscillations
could be severely diminished by a non-linear parametric resonance that could bleed energy
into very short-wavelength modes, dissipating this g-mode oscillation thermally.
Such daughter modes are too small to simulate with current grids, but the weakness
of the consequent explosions and the possibility of an important additional damping mechanism
make this ``acoustic" solution sub-optimal.  Moreover, if the neutrino mechanism in any form obtains,
it would naturally abort the acoustic mechanism.

\begin{figure}
\centerline{
\includegraphics[height=.80\textheight,angle=0]{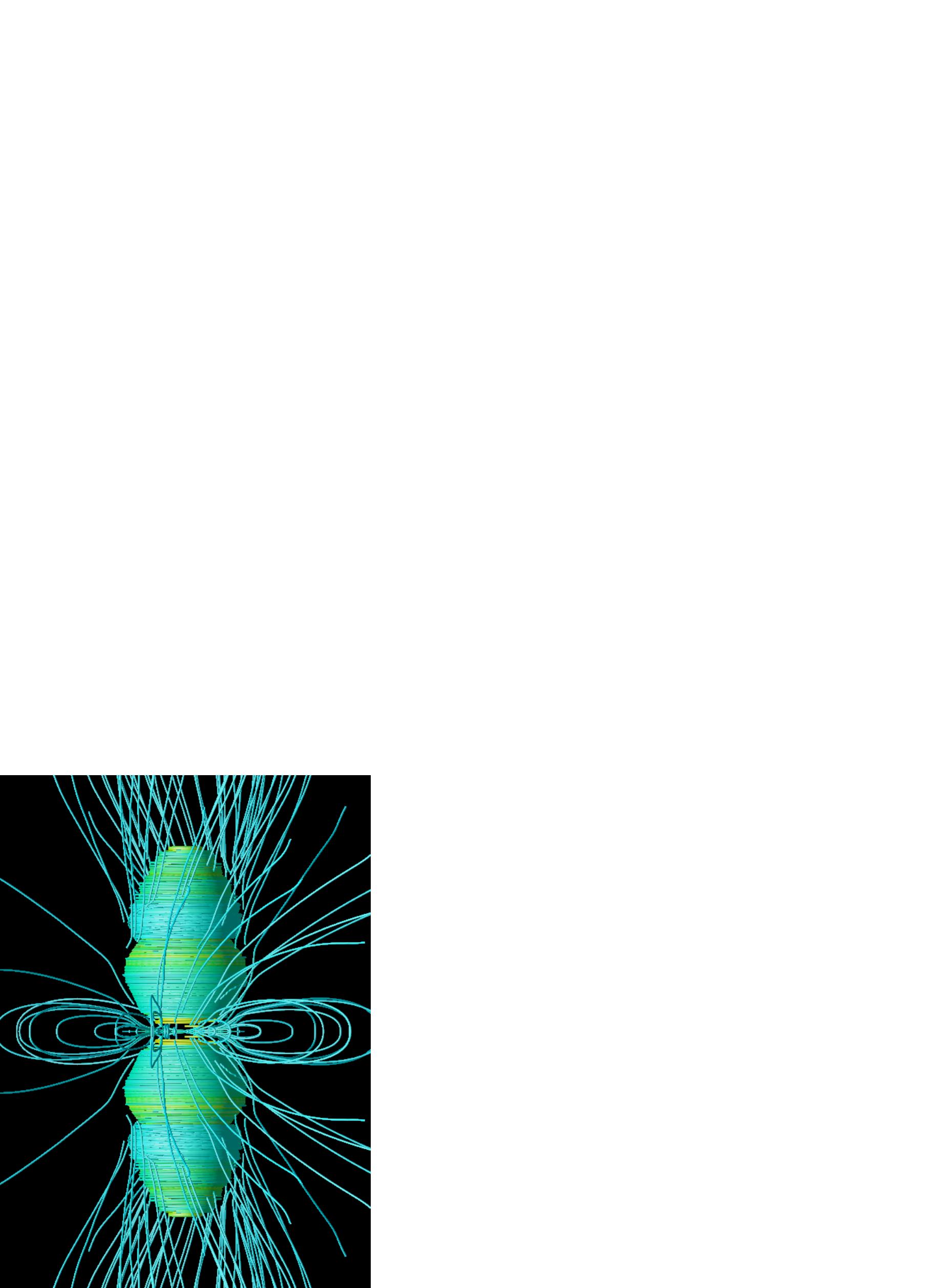}
}
\caption{An early snapshot during the bipolar explosion of the mantle of a rapidly-rotating 
core.  Depicted are representative magnetic field lines.  The scale is 2000 km from top to bottom.
The extremely twisted lines are just interior to the shock wave being driven out by magnetic pressure generated
within 200 milliseconds of bounce in a ``2.5"-dimensional magneto-radiation-hydrodynamic 
calculation conducted by Burrows et al. (2007d).  The progenitor employed for this calculation 
was the 15-M$_{\odot}$ model of Heger et al. (2005).
\label{mhdfig}
}
\end{figure}

Rapid rotation with magnetic fields should naturally lead to vigorous explosions
and this MHD mechanism has a long history (LeBlanc \& Wilson 1970; Symbalisty 1984; 
Akiyama et al. 2003; Akiyama \& Wheeler 2005).
The free energy of differential rotation available at bounce if the progenitor core
is rapidly rotating ($P_i \sim 1-2$ seconds)\footnote{translating into a core rotating 
with $\sim$1.5- to 4-millisecond periods} is a potent resource, naturally channeled into bipolar
jet-like explosions (Burrows et al. 2007d). Figure \ref{mhdfig} portrays such
a magnetic explosion.  A variant of this mechanism has been suggested
for gamma-ray bursts (MacFadyen \& Woosley 1999).  However, pulsar spin data
indicate that the generic progenitor core spin rates must be rather low, and that
at most $\sim$1\% of collapses can be via such an MHD mechanism. Hence, MHD-driven
explosions can't be the generic core-collapse supernova channel. However, the neutrino heating 
mechanism may not be able to provide explosion energies above a few Bethes, and this suggests that
the best explanation for the rare energetic ``hypernovae" might be MHD power
due to very rapidly rotating cores.

There are numerous ambiguities and problems with the current generation of 2D simulations 
that are not reproducing the core-collapse supernova phenomenon.  One symptom (or feature) 
of the problem in 2D may be the long time to explosion seen  
by, e.g., Marek \& Janka (2009), Suwa, et al. (2010), M\"uller, Janka, \& Marek (2012), and Burrows et al. (2006).
Waiting $\sim$400$-$1000 milliseconds to explosion may, certainly in the case of the 
neutrino-heating mechanism, ``waste" the neutrinos emitted in the interval after bounce.
However, Nature is not 2D and it could be that an impediment to progress in 
supernova theory over the last few decades is a lack of access to codes, computers, 
and resources with which to properly simulate the collapse phenomenon in 3D. This
could explain the agonizingly slow march since the 1960's towards demonstrating a robust
mechanism of explosion. The difference in the character of 3D turbulence, with its extra
degree of freedom and energy cascade to smaller turbulent scales than are found in 2D,
might relax the critical condition for explosion (Burrows \& Goshy 1993).
Indeed, Murphy \& Burrows (2008) have shown that the critical condition is more
easily met in 2D than in 1D and, recently, Nordhaus, Burrows, Almgren, \& Bell (2010)
conducted a similar parameter study comparing 1D, 2D, and 3D and found that going
from 2D to 3D could lower the threshold for explosion still more.  However, 
Hanke et al. (2012) have called this conclusion into question.  Nevertheless, 
these developments suggest that it is a few tens of percent  
easier to explode in 3D than in 1D, and that 
full 3D simulations, but with competitive multi-group neutrino transport, 
might be needed to properly address this long-standing problem in computational nuclear 
astrophysics. Such ``heroic" 3D simulations will be very computationally challenging, 
but are the future (Takiwaki, Kotake, \& Suwa 2012). 

\section{Important Features of Supernova Theory}
\label{features}

Following my description of the core-collapse scenario and 
my brief summary of the current status of the numerical theory, I now embark
upon discussions of select topics, that though important, are often ignored, 
assumed, or misrepresented. However, I attempt merely to provide simple, yet 
useful, insights into basic supernova physics $-$ rigor is not here my goal. In 
the process, I highlight some of the central themes and myths of 
core-collapse supernova theory. Since I conduct this survey in lieu of a final 
resolution of the supernova problem, the reader is encouraged to retain an 
open mind, and forgiven for retaining a critical one. 

\subsection{General Themes}
\label{themes}

Frequently missing in general discussions of core-collapse supernovae is that they are
gravitationally-powered.  Nuclear burning during explosive nucleosynthesis of the outer mantle
after the explosion is well along might contribute at most $\sim$10\% of the blast energy.  
A full solar mass of oxygen and/or carbon would have to burn to iron peak to yield 
one Bethe. Given all extant progenitor model profiles, much less than that amount of fuel is 
close enough to the core to achieve by shock heating sufficient temperatures ($\ge$$4 \times 10^{9}$ K). 
Moreover, before explosion, any infalling fuel will be burned uselessly during collapse, and the ashes 
will be dissociated by the stalled shock and then buried in the core.  
The neutrino energy emitted from the core and absorbed in the proto-neutron star mantle that is required for 
the neutrino-powered model has its origin in compressive work on the matter of the core 
by gravitational forces.  The trapped leptons are compressed and the matter is heated to high 
temperatures and thermal energies, both of which represent stored energy eventually to be radiated.  
Rotationally-powered and magnetic models ultimately derive their energy from the conversion 
of gravitational binding energy changes during implosion into rotational kinetic energy 
(roughly conserving angular momentum), and then into magnetic energy.  

One of the characteristics of core-collapse supernovae that distinguishes them from thermonuclear 
(Type Ia) supernovae is that they leave a residue, the neutron star or black hole.  It is not necessary
to disassemble and unbound this remnant to infinity, thereby paying a severe gravitational binding 
energy penalty. Since neutrino radiation renders the PNS more and more bound with time, if explosion were to require
complete disassembly to infinity, time would not be on the side of explosion.  If fact, core-collapse supernovae
would probably not be possible.  However, in a fundamental sense, core-collapse supernova 
involve the transfer of energy from the core to the mantle, leaving the core behind. It is the 
mantle that is ejected.  This mantle may start near $\sim$100 km, not the canonical radius of $\sim$10 km,
and hence is much less bound.  All CCSN explosion models are different models for core-mantle
energy transfer, be it direct hydrodynamic (core piston), neutrino (mantle heating by core neutrinos), 
or MHD (mantle B-field amplification, tapping core rotation). One can bury in the residue a 
binding energy problem that could have gotten progressively worse with time.

It may seem curious that the average thermonuclear supernova has an explosion energy 
that is similar to that of the average CCSN. However, the energy for the former derives from the burning 
of a large fraction of something like a Chandrasekhar mass, while an energy bound for the latter might be set 
in part by the gravitational binding energy of the stellar mantle surrounding a Chandrasekhar mass residue. 
The core-collapse explosion must eject this bound mantle. Burning yields $\sim$0.5 MeV per baryon,
and the binding energy of a Chandrasekhar mass is roughly $m_ec^2$ per electron, where $m_e$ is the electron mass.
The latter obtains due to the fact that the Chandrasekhar mass is defined by the onset of relativity
for the majority of its electrons.  Since Y$_e$ is $\sim$0.5, and perhaps 50\% of a C/O white dwarf burns
to make a Type Ia supernova, both total energies are very approximately the number of baryons in a 
Chandrasekhar mass times $\sim$0.5$\times$0.5 MeV.  This is very approximately one Bethe.  The binding energy 
of the white dwarf core of a massive progenitor and the binding energy of the stellar envelope
around it will crudely scale with one another, due to the pseudo-power-law density profile of the latter.
Therefore, the energy scales for both thermonuclear and gravitational supernovae explosions 
(the latter in the sense of a bound) are of comparable magnitude.  This argument may be 
good to a factor of a few, but that it is good at all in a Universe with a much wider 
potential range of energies is perhaps noteworthy.

\subsubsection{Eigenvalue Problems}
\label{eigen}

\begin{figure}
\centerline{
\includegraphics[height=.80\textheight,angle=0]{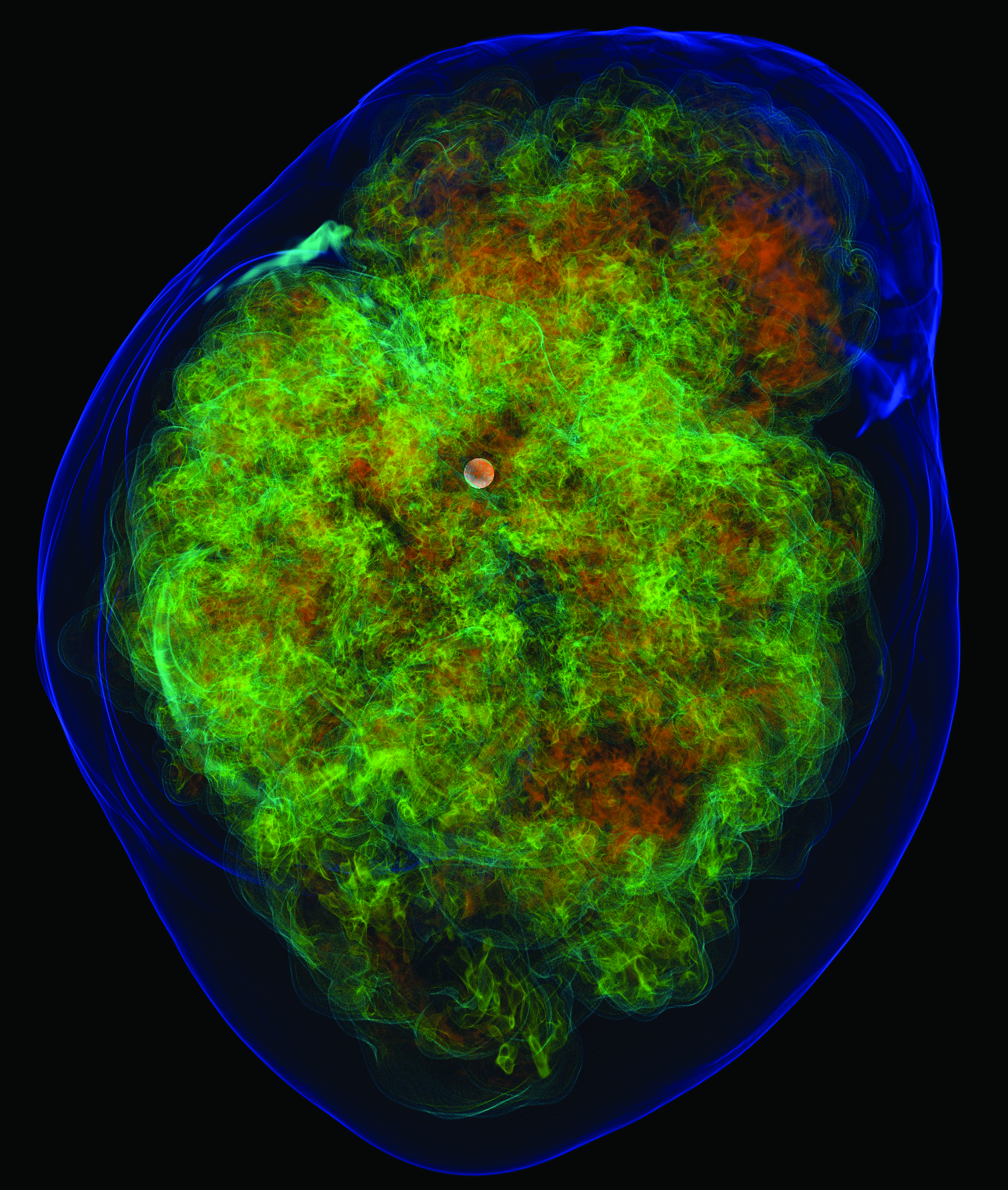}
}
\caption{The debris field generated in a 3D neutrino-driven explosion approximately 200
milliseconds after its onset.  The scale from top to bottom is 1000 km. The blue exterior
is a rendering of the shock wave, the colored interior is a volume-rendering of the entropy
of the ejecta, and the sphere in the center is the newly-born neutron star. (Numerical data taken
from Nordhaus et al. (2010) and the 15-M$_{\odot}$ progenitor from Woosley \& Weaver 1995 was used.)
\label{chupa14}
}
\end{figure}

There are two important approximate eigenvalue problems associated with core collapse.  
The first involves the post-bounce, pre-explosion PNS structure, bounded by an accretion 
shock. The hydrodynamics during this phase is roughly quasistatic, so one can drop the 
time derivatives to arrive at a set of simultaneous ordinary differential equations for 
the hydrodynamic profiles interior to the accretion shock. With shock outer boundary 
conditions, an inner core mass ($M_c$), a given accretion rate ($\dot{M}$), and 
given core neutrino luminosities ($L_i$), one can convert this into an eigenvalue problem
for the radius of the shock ($R_s$).  One derives $R_s$ in terms of the control parameters
$\dot{M}$, $L_i$, and $M_c$, given assumptions about mantle heating due to $L_i$ and cooling
due to electron and positron capture on nucleons. As Burrows \& Goshy (1993) showed, there
is a critical curve in $L_i$ versus $\dot{M}$ space (ceteris paribus) above which there are no solutions
to this eigenvalue problem.  As one increases $L_i$ for a given $\dot{M}$, $R_s$ increases, but it
can not increase to arbitrary values. At a critical curve $L_i$, $R_s$ is finite, but above the
critical value of $L_i$ for a given $\dot{M}$ the steady-state problem does not have 
a solution. The absence of a solution can be considered an approximate condition for explosion by 
the neutrino heating mechanism. The subsequent evolution is dynamical, with continued 
neutrino heating depositing energy to power the explosion and expansion of the gas lowering 
the temperature (not the entropy!) and, thereby, the cooling rates. Furthermore, as the mantle 
accelerates into explosion, the matter recombines from nucleons and alpha particles into  
iron peak nuclei (with $(Z,A)$ depending upon Y$_e$), thereby ``returning" to the expanding
matter the nuclear binding energy of the ejecta ``lost" to shock photodissociation. 
These ``original" ejecta may not contain more than a few hundredths of a solar mass, but since 
$\sim$8$-$9 MeV are liberated per baryon, only $\sim$0.1 solar masses of ejecta are needed to 
supply approximately one Bethe to the explosion.  
Similar critical curves can be derived that include multi-dimensional turbulence in 2D 
(Murphy \& Burrows 2008) and 3D (Nordhaus et al. 2010) and these seem to be lower, 
facilitating explosion. Figure \ref{chupa14} depicts the debris field of such 
a 3D explosion. However, the detailed reasons for this dimensional boost are 
still being studied. 

The second eigenvalue problem presents itself after explosion and is the neutrino-driven wind
that emerges from the proto-neutron star.  An old model (Parker 1958) for the solar wind started by
assuming that the plasma above the solar photosphere was in hydrostatic equilibrium and that
the energy luminosity due to electron conduction through this atmosphere was constant. 
Since electron conductivity in a hot plasma depends almost solely on temperature ($\propto T^{5/2}$),
one derives temperature as a function of radius ($T \propto 1/r^{2/7}$).  Hydrostatic equilibrium of an ideal gas
can then be integrated to derive the pressure as a function of radius.  What one finds is that 
around a spherical star the pressure must be finite at infinity!  This shows that
in order to maintain hydrostatic equilibrium the star must be artifically embedded in a high pressure gas.
Since the pressure in the interstellar medium is very low, this atmosphere cannot be stable and it
would spontaneously erupt as a wind.  The flow would transform into a steady-state outflow,
with a sonic point and a supersonic asymptotic speed.  Though we now know this 
particular physical model does not apply to the Sun, these arguments led to the prediction 
of the existence of the solar wind. The wind mass loss rate would be a function of the 
driving luminosity, the stellar mass, and details of the heating process.  
It can be shown that if the derived temperature profile falls off more slowly than $r^{-1}$,
such an atmosphere is similarly unstable.  

The relevance of this scientific parable is that the same arguments can (more legitimately)
be applied to the mantle of the PNS.  The balance of neutrino heating ($\propto 1/r^2$)
and neutrino cooling ($\propto T^6$) yields $T \propto 1/r^{1/3}$, with a power-law index 
less than one.  Therefore, without a bounding pressure the PNS atmosphere is unstable
to a neutrino-driven wind (Duncan, Shapiro, \& Wasserman 1986; Burrows 1987; Burrows, 
Hayes, \& Fryxell 1995).  In the core-collapse context this bounding pressure 
is provided by the accretion ram and, while the shock is stalled, the wind 
is thwarted.  However, after explosion and after the pressure around
the PNS has subsided due to the progress of the supernova explosion, 
a neutrino-driven wind naturally emerges. The eigenvalue problem for $\dot{M}$ as a 
function of driving luminosity and PNS mass is easily solved.

Therefore, in the context of the delayed neutrino heating model, the supernova itself 
is the dynamical transitional state between two quasi-steady-state eigenvalue problems, one accretion and 
the other a wind. The mechanical power in the wind is lower than the instantaneous 
power being poured into the early supernova because the absorbing mass and neutrino optical 
depth ($\tau_{\nu}$) of the atmosphere above the PNS around the gain region are much larger than 
in a tenuous wind.  At the onset of explosion, how much larger the absorbing mass 
and depth are determines how much power is available for explosion. If $\tau_{\nu}$ is large when 
the $\nu_e$ and $\bar{\nu}_e$ luminosities are large, their product will be large
and the explosion will be robust. The simulated explosions of the 8.8-M$_{\odot}$
progenitor model of Nomoto \& Hashimoto (1988), with the very steep density gradient outside the core,
transitioned so quickly into a wind that these model supernovae were effectively wind-powered
and had the correspondingly low explosion energy alluded to earlier (Kitaura et al. 2006; 
Burrows et al. 2007c). 

\subsubsection{Simultaneous Accretion and Explosion}
\label{simul}

When breaking spherical symmetry in the context of the multi-dimensional 
instabilities seen in modern supernova simulations, a feature (some would say a virtue) 
of many proposed core-collapse supernova mechanisms is that during the early 
phases of explosion there can be simultaneous explosion and accretion 
(Burrows et al. 2007a). Continued accretion onto the PNS from one direction 
can supplement the energy available to power explosion in another. 

The neutrino mechanism, in part powered by accretion luminosity, is a good example.  
In spherical symmetry, explosion is the ``opposite" of accretion, and that source 
of neutrino driving subsides early after the onset of explosion. However, 
if the symmetry is broken and a neutrino-driven explosion first occurs 
in one direction, continued accretion onto the PNS from another direction 
can help maintain the driving neutrino luminosities. Though such accretion 
might be restricted to a small quadrant, neutrino emissions are always 
much more isotropic than matter distributions (Ott et al. 2008), with the 
result that accretion almost anywhere on a PNS surface leads to emitted neutrinos almost 
everywhere. In a sense, the same is true for the MHD mechanism, wherein explosion is
bipolar along the rotation axis, while the spinning PNS accretes along 
the equator. Conserving angular momentum, such accreta continue to bring in the 
kinetic energy of differential motion needed to maintain the magnetic energy and pressure
that power the explosion. As long as equatorial accretion continues, the core is an ``engine"
with a power source.  After equatorial accretion ceases and the explosion assumes a more
isotropic distribution, the engine subsides, but the supernova (or hypernova) is launched. 
The acoustic mechanism is the quintessential process that exploits simultaneous accretion
on one side, which maintains the driving core g-mode oscillations, to power an explosion
in the other direction (Burrows et al. 2006,2007b).

\subsubsection{Energetics}
\label{energy}

Determining the energy of a detailed numerical explosion can be more awkward 
than one might think.  Usually limited by the small size of the computational domain 
(e.g., 5000$-$20000 km), a successful shock wave encounters this border and perforce stops 
within hundreds of milliseconds of the start of explosion and long before the explosion 
energy has asymptoted.  For such calculations, neutrino energy deposition is still 
ongoing, recombination of the nucleons and alphas has not completed, and, importantly,
the baryon mass cut between the final PNS and the ejecta is not determined.  
In fact, the mass cut has never been consistently determined for any detailed 
numerical core-collapse supernova model. In addition, the explosion must work 
to unbind the star exterior to the mass cut, and this matter (most of the remaining star)
can be bound by a few Bethes (see Figure \ref{bind}).  The larger this binding energy, the
stronger the explosion in the core needs to be to achieve a given final ejecta supernova 
kinetic energy. The large magnitude of this binding energy for the most massive progenitors
may be instrumental in either aborting what may have started as a promising supernova,
or in ensuring that a black hole, rather than a neutron star, remains.  In fact, the 
large binding energies for matter exterior to a canonical 1.5 M$_{\odot}$ in
progenitors that one has in the past thought should supernova and leave neutron stars (such 
as 20 or 25 M$_{\odot}$ ZAMS stars) suggest that either the core explosion must 
be very vigorous or that explosions in such stars fizzle.  This would be unfortunate,
since it is thought that the more massive progenitors are the likely primary 
sources for the oxygen that we see in abundance in the Universe. However, the relevant 
outer binding energies are functions not only of progenitor mass, but of modeler. This 
is yet another indication of the centrality of progenitor models to our understanding 
of the outcome of collapse.

In the neutrino heating model, one way to achieve higher explosion energies may be to explode early.
In the current paradigm, during the delay to explosion the neutrino energy deposited is 
reradiated and useless.  There is no accumulation of energy in the post-shock
mantle until the explosion is underway.  This suggests that a long delay to explosion
may be detrimental, wasting as it does the neutrinos radiated by the core before the explosion 
commences. However, an early explosion may be correspondingly useful, with perhaps some later fallback. 
Such an early onset may be easier in 3D (Nordhaus et al. 2010). In addition, and 
counter-intuitively, both before and at the onset of explosion, the enthalpy fluxes 
and PNS mantle energies are negative, the latter at times even when the recombination
energy is accounted for. For all the viable explosion mechanisms, the supernova does 
not attain its final energy at the instant of explosive instability, but must be 
driven after it starts and still needs to overcome the PNS and outer stellar envelope gravitational 
binding energies. Currently, no detailed neutrino-driving simulation has come within 
an order of magnitude of achieving this requirement, except perhaps the singular 
8.8-M$_{\odot}$ neutrino-wind-driven model.

\subsubsection{Conditions for Explosion by the Neutrino Mechanism}
\label{cond}

The neutrino mechanism, legitimately the front-runner in the 
CCSN mechanism sweepstakes, has engendered much speculation concerning 
the physical conditions for explosion.  In section \ref{eigen}, I 
described the critical condition between $\dot{M}$, $L_i$, and $M_c$ 
that signals instability to explosion.  This condition, 
with corrections to account for multi-D effects, still seems close to
capturing the essence of the explosion condition.  A detailed 
perturbation analysis of these steady states to explore the 
complex eigenfrequencies of the monopolar and low-order pulsational modes,
in particular to determine when their imaginary parts change sign, 
would lend a useful additional perspective (Yamasaki \& Yamada 2007). 

However, other explosion conditions have been proffered which aid understanding 
(e.g., Pejcha \& Thompson 2012; Burrows 1987; Janka 2001). By and large, all these are roughly equivalent.
All sensible conditions must recognize that since the matter interior 
to the pre-explosion shock is in sonic contact the condition
for explosion must be a global one. A local condition, say on the pressure at the shock,
has little meaning and can be misleading. It is the entire mantle structure 
that is exploding.  Moreover, the discussion in section \ref{energy} 
indicates how subtle things might be, with explosion commencing even when 
various otherwise obvious quantities associated with energy or energy
flux are negative.

It should be noted that the gain region interior to the shock, in which
neutrino heating outpaces capture cooling, surrounds a net cooling region 
where cooling dominates.  The inner boundary of the cooling region coincides, 
more or less, with the $\nu_e$ and $\bar{\nu}_e$ neutrinospheres.  This region 
gradually sinks in due to energy and lepton loss, steadily sending out rarefaction 
waves that are partially responsible for undermining the gradual outward progress 
of the quasi-static shock wave and the growth of the gain region. The larger 
the gain region and the smaller the cooling region the more likely the 
mantle is to explode.  However, the cooling power generally outstrips net 
heating and this fact is one of the primary impediments to explosion.  If there
were no cooling region, or if the cooling power in the cooling region were significantly
reduced, neutrino absorption would quite easily lead to explosion.  A focussed study 
on how Nature might accomplish this might bear fruit.

A useful approximate condition for explosion is when the 
characteristic neutrino heating time in the gain region ($\tau_{H}$) 
exceeds the advection time ($\tau_{adv}$) through it (Thompson et al. 2005; Janka 2001). 
For every set of definitions of these particular times, and there are a variety of definitions
which can vary by factors (Murphy \& Burrows 2008), the critical ratio itself should 
be calibrated using hydro.  In any case, $\tau_{adv}$ can be set equal to $\Delta r/v_{eff}$,
where $v_{eff}$ is some effective speed through the gain region that incorporates
the sinuous trajectories of Lagrangian particles in multi-D.  Quite naturally, $\tau_{adv}$
is larger in 3D than in 2D, and larger in 2D than in 1D (spherical). $\tau_{adv}$ can also
be written as $\Delta M/\dot{M}$, where $\dot{M}$ is the accretion rate through the shock
and $\Delta M$ is the mass in the gain region.

$\tau_{H}$ can be defined as the internal energy in the gain region divided by the neutrino
heating power.  The latter is approximately $L_{\nu_e} \tau_{\nu}$, where, again, $\tau_{\nu}$
is the electron neutrino optical depth. Therefore, setting $\tau_{H}$ equal
to $\tau_{adv}$ gives us $L_{\nu_e} \sim (\varepsilon/\tau_{\nu}) \dot{M}$, where $\varepsilon$
is the specific energy in the gain region.  $\varepsilon$ might scale with the escape speed 
squared ($v_{esc}^2$) at the shock, and this quantity scales with the core mass and the inverse 
of the shock radius (i.e., $v_{esc}^2 \sim 2GM_c/R_s$). This rough relation yields a critical 
$L_{\nu_e}$ versus $\dot{M}$ curve with a slope of ($\varepsilon/\tau_{\nu}$), which itself may 
be a weak function of $L_{\nu_e}$, $\dot{M}$, and $M_c$ that is better calculated numerically.  
However, this relation can be recast by noting that $\tau_{\nu} \sim \kappa_{\nu_e} \rho \Delta r$, 
where $\kappa_{\nu_e}$ is the electron neutrino absorption opacity and $\rho$ is some mean 
mass density in the gain region.  The result is
\begin{equation}
L_{\nu_e}(crit) \sim \frac{4\pi G\dot{M}}{\kappa_{\nu_e}}\frac{M_c}{\Delta M}{R_s}\, .
\label{crit}
\end{equation}
The actual constant of proportionality will depend upon details.  Nevertheless, equation 
(\ref{crit}) states that the higher the absorptive opacity or the ratio of the mass in the gain region
to the core mass the lower the critical luminosity for a given $\dot{M}$ and ${R_s}$.  However, the quantity 
${R_s}/{\Delta M}$ varies slowly with the control parameters $L_{\nu_e}$, $\dot{M}$, and $M_c$, making equation(\ref{crit})
a more direct connection between them that succinctly summarizes the critical 
condition of Burrows \& Goshy (1993). A correction factor to account for multi-D 
effects could be added in the denominator.  Not surprisingly, the critical curve 
and the $\tau_{H} = \tau_{adv}$ condition are roughly equivalent.

\subsubsection{Instability to Finite Perturbation}
\label{nonlinear}

The stalled accretion shock becomes unstable to outward expansion and explosion when (or near when) 
the critical $L_{\nu_e}$-$\dot{M}$-$M_c$ condition is met and exceeded.  However, it is also unstable
to an abrupt, finite jump in its position.  If, by some mechanism, the shock were to be jolted suddenly
to larger radii, the consequently lower matter temperatures would transiently result in a much lower 
integrated neutrino cooling rate behind the shock, while the larger radius of the shock would result 
in a larger gain region.  Such a sudden, favorable, and finite shift to more heating and less 
cooling (and to a lower accretion ram pressure at the shock) could be irreversible and an explosion 
might be ignited. However, the magnitude of the necessary finite perturbation is not known.
The agency of such a jolt is also not known, but the accretion of density discontinuities
in the progenitor at composition and entropy boundaries (such as the inner edges of the 
silicon- or oxygen-burning shells) is seen in hydrodynamical simulations to result in a quick 
outward (though modest) excursion in the average shock radius. If the actual density jumps are larger than
in the current generation of pre-collapse models, or if there are significant variations 
in the density or velocity profiles of the post-bounce accreting matter, the necessary finite 
perturbations may be available.  To be sure, this discussion is highly speculative, but the 
possibility remains intriguing that the mantle shell interior to the stalled shock could be 
nonlinearly unstable to explosion by large, finite-amplitude perturbations.

\subsection{Persistent Myths}
\label{myths}
 
There are a number of what I would call ``myths" that have 
emerged and persisted, despite compelling physical counter-arguments.
To be sure, my list is idiosyncratic and the list of others may 
be different. One myth is that neutrino transport is more difficult
than photon transport, requiring only a specialist's touch.
In fact, non-LTE photon transport, with its multitude of level populations,
spectroscopic data, collisional processes, and lines, is much more difficult 
and challenging than neutrino transport. The latter involves only 
continuum opacities and emissivities and one rate equation, that 
for Y$_e$, which is coupled only to $\nu_e$ and $\bar{\nu_e}$
transport.  True, there are six neutrino species and one does require
knowledge of neutrino-matter couplings.  Also, some astronomers
shy away from the nuclear and neutrino realm, and are more interested in
the dominant signatures of the Universe $-$ those in photons.  However, the 
physics of neutrino transport is far more straightforward than the physics of 
atomic and molecular spectroscopy and of the myriad collisional processes 
in a heterogeneous soup of elements and ions. In addition, the numerical art of photon 
transport is a rigorous, well-developed, subject with expertise spread 
around the world, whereas experts in neutrino transport are few and 
far between.  

Another myth is that since $\sim$$3 \times 10^{53}$ ergs of binding energy
is emitted during the long-term ($\sim$10-50 second) PNS cooling 
and deleptonization phase, whereas the average core-collapse supernova 
involves only $\sim$$10^{51}$ ergs, the CCSN is a less than 1\% affair, 
requiring exquisite precision and numerical care in approaching its theory. 
In fact, since the $\nu_e$ and $\bar{\nu_e}$ absorption optical depths in the gain region 
are $\sim$4$-$10\% and this is the fraction of the core $\nu_e$ and $\bar{\nu_e}$ 
luminosities absorbed in the mantle, the core-collapse neutrino mechanism is 
more like a $\sim$4$-$10\% affair. Nevertheless, it is often suggested that 
every detail makes a difference to the overall outcome, as if the mechanism 
itself hinged upon them. The result has been that minor effects have sometimes 
been allowed to loom large, often confusing those not intimating involved 
in the research.  Examples are $\nu$-$\bar{\nu}$ annihilation, 
neutrino-electron scattering, electron capture on infall, neutrino-neutrino 
oscillations, and the nuclear symmetry energy, to name only a few.  
This is not to say that all these topics are not to be addressed, nor 
that the ultimate theory can afford to ignore them.  Incorporating the 
correct physics and performing detailed simulations will certainly be necessary
to obtain the correct numbers.  However, when the best, most detailed, extant exploding 2D CCSN 
simulations may not be reproducing observed supernova energies by an order of magnitude 
perhaps a focus on details at the expense of global understanding 
is unfounded.  Something much more important may 
be missing.  


A more innocuous myth is that a stellar-mass black hole can form directly.
In fact, in the context of the collapse of an effective-Chandrasekhar-mass core,
the inner homologous core will always rebound into the outer core, generating 
a shock wave.  Interior to this shock wave at its inception and during its 
early life is only $\sim$1.2$-$1.5 M$_{\odot}$ of material and this is not enough 
for the core collectively to experience the general-relativistic instability that 
leads to stellar-mass black holes. Importantly, the inner core and shocked mantle are out 
of sonic contact with the supersonically infalling outer core, and, hence, do not yet
``know" whether enough mass will accumulate to transition to a black hole. Sufficient 
matter must accrete through the shock before the core can go unstable. The wait, during which
the core will fatten, might require hundreds of milliseconds to seconds. after which a second dynamical collapse to a black hole 
will ensue. Therefore, black hole formation is always preceded by an intermediate PNS stage and
cannot proceed directly (Burrows 1984; Sumiyoshi, Yamada, \& Suzuki 2007ab; Fischer et at. 2009; O'Connor \& Ott 2011). 

Neutrino heating in the gain region naturally generates a negative entropy 
gradient in the steady state.  This gradient is unstable to convective overturn
(the generalized Rayleigh-Taylor instability) and leads to turbulence. This turbulence
and the corresponding aspherical flow patterns, with neutrino-driven upflowing plumes and associated 
downflows, has been seen in multi-D simulations since the 1990's and are central features 
of supernova theory.  

\begin{figure}
\centerline{
\includegraphics[height=.70\textheight,angle=0]{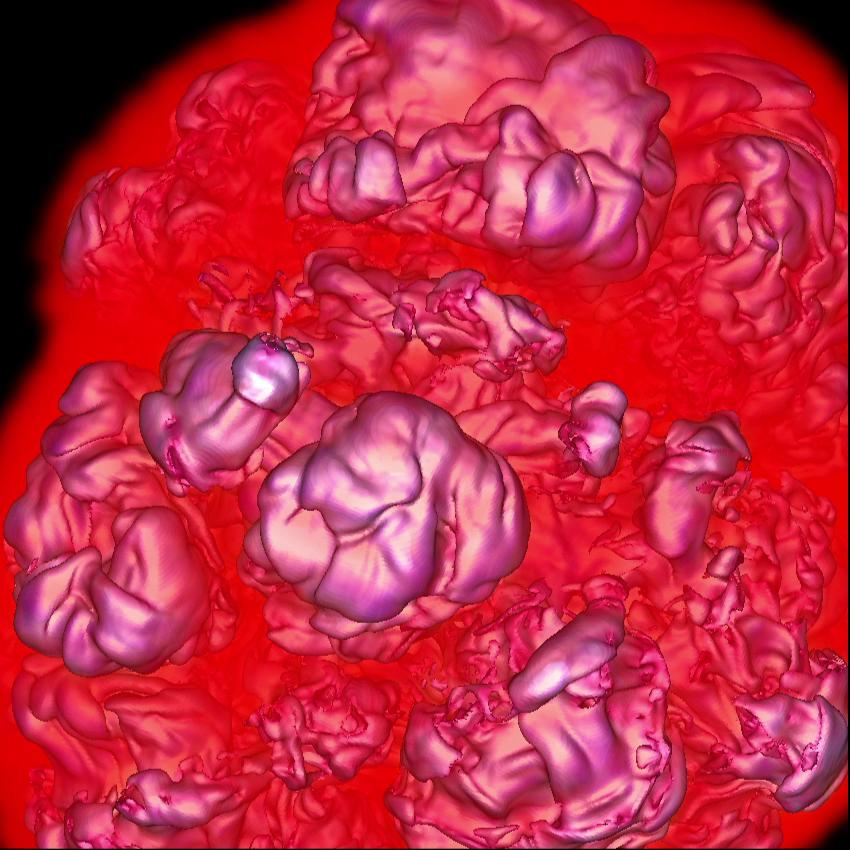}
}
\caption{A different rendering of the 3D neutrino-driven explosion shown in Figure \ref{chupa14}, but 
approximately 25 milliseconds after its onset.  The scale from top to bottom is $\sim$500 km. The prominent
bubble structures in magenta are isoentropy surfaces and are provided to highlight the neutrino-heated 
bubbles that seem to be generic in current 3D simulations. The bounding shock is not shown. The red colored 
interior (between the bubbles) is an experimental (only partially successful) volume-rendering of the density. (Numerical data taken
from Nordhaus et al. (2010) and the 15-M$_{\odot}$ progenitor from Woosley \& Weaver 1995 was used.)
\label{bubble}
}

\end{figure}

Later, Blondin, Mezzacappa, \& DeMarino (2003) identified an instability of the stalled 
shock, the standing accretion shock instability (``SASI"), that has been well-diagnosed 
and studied by Foglizzo (2002,2009), Blondin \& Mezzacappa (2006), Foglizzo, Scheck, 
\& Janka (2006), Foglizzo et al. (2007), Yamasaki \& Foglizzo (2008), Scheck et al. 
(2008), Iwakami et al. (2008), Kotake et al. (2009), and Foglizzo et al. (2012).  
%
%
In axisymmetric (2D) studies of the pure SASI (without neutrinos), many 
witnessed a vigorous dipolar ($\ell = 1$) ``sloshing" mode that superficially seems like the corresponding
sloshing motion seen in full 2D radiation/hydro simulations.  This has led many, I believe 
incorrectly, to associate the motion seen in most full neutrino-transport runs with the 
motions seen in the simplified neutrino-free studies. Moreover, since the energy cascade in 2D
turbulence (neutrino-driven or otherwise) is ``backwards" (Boffetta \& 
Musacchio 2012), from small to large scales, and the SASI instability, by its nature, 
is on large scales (small spherical harmonic index $\ell$), this chance correspondence 
of dominant scales in 2D may also in part explain the confusion. More importantly, preliminary 
calculations performed in 3D (e.g., Burrows, Dolence, \& Murphy 2012; Dolence, Burrows, \& Murphy 
2012; Hanke et al. 2012) do not show the ``sloshing" dipolar motion along 
an axis many have come to associate with the SASI and that some have suggested is 
crucial to the CCSN mechanism (e.g., Marek \& Janka 2009; Hanke et al. 2012). What are seen
are bubble structures and plumes indicative of buoyant convection (Figure \ref{bubble}).  Hence, the prominent
$\ell = 1$ ``SASI" mode may be an artifact of 2D and its inverse cascade.  However, when 3D simulations
with reasonable physics become more readily available, the fact that the cascade in 3D 
is in the opposite direction, but the dominant SASI modes are still on large scales, 
should help clarify the true nature of the turbulence seen.  

I include this discussion on the SASI under ``myths" because 1) current 3D simulations
do not show the signature features of the SASI ostensibly seen in 2D, and 2) I fully expect that
when credible and self-consistent 3D radiation/hydrodynamic simulations are performed
an objective analysis of its role will reveal it often to be subordinate to neutrino-heated
buoyant bubble convection.  The reader is cautioned that not every supernova researcher 
will agree with this characterization, so caveat lector. Nevertheless, and curiously, though simulations
with neutrino transport almost always show the classic rising bubble and downflow patterns of buoyancy-driven
non-linear convection, many people, even practitioners, had started to refer to all turbulent 
motions behind the shock as the ``SASI." This may have served to confuse both insiders and outsiders alike.
In summary, I am led to suggest that if neutrino driving is the energetic agency of explosion, it is naturally also the primary
agency of the turbulence that aids explosion.

\section{Conclusions}
\label{conclusions}

I have sought in this short review of core-collapse supernova theory to clarify its major facets,
lay bear its physical underpinnings, and briefly summarize its current status, as I see it.  Furthermore, I have 
tried to identify some of what I consider to be myths that have crept into theoretical discourse.
However, this colloquium is a personal view of the theoretical landscape. There is clearly much more to be done 
before a cogent explanation of this central astrophysical phenomenon is available and verified, and some
of what I have suggested here may not survive future scrutiny. Nevertheless, it is hoped that at the 
very least this paper provides some novel and useful insights into the physics and astrophysics 
of both core collapse and supernova explosions.

\begin{acknowledgments}
The author acknowledges fruitful collaborations with, conversations with,
or input from Josh Dolence, Jeremiah Murphy, Christian Ott, Jeremy Goodman, Louis Howell, Rodrigo Fernandez,
Manou Rantsiou, Tim Brandt, Eli Livne, Luc Dessart, Todd Thompson, Rolf Walder,
Stan Woosley, Ann Almgren, John Bell, and Thomas Janka. He would also like to thank Hank Childs and the VACET/VisIt
Visualization team(s) and Mike Chupa of PICSciE for help with graphics and with developing 
multi-dimensional analysis tools.  A.B. is supported by the Scientific Discovery through
Advanced Computing (SciDAC) program of the DOE, under grant number DE-FG02-08ER41544,
the NSF under the subaward no. ND201387 to the Joint Institute for Nuclear Astrophysics (JINA, NSF PHY-0822648),
and the NSF PetaApps program, under award OCI-0905046 via a subaward
no. 44592 from Louisiana State University to Princeton University.
Some of the author's science employed computational resources provided by the TIGRESS
high performance computer center at Princeton University, which is jointly supported by the Princeton
Institute for Computational Science and Engineering (PICSciE) and the Princeton University Office of
Information Technology; by the National Energy Research Scientific Computing Center
(NERSC), which is supported by the Office of Science of the US Department of
Energy under contract DE-AC03-76SF00098; and on the Kraken supercomputer,
hosted at NICS and provided by the National Science Foundation through
the TeraGrid Advanced Support Program under grant number TG-AST100001. 
\end{acknowledgments}


\onecolumngrid

\def\apj{Astrophys.~J.\ }
\def\apjl{Astrophys.~J.~Lett.\ }
\def\aj{Astron.~J.\ }
\def\mnras{Mon.~Not.~R.~Astron.~Soc.\ }
\def\ann{Annu.~Rev.~Astron.~Astrophys.\ }
\def\araa{Annu.~Rev.~Astron.~Astrophys.\ }
\def\apjs{Astrophys.~J.~Suppl.\ }
\def\pasp{Publ.~Astron.~Soc.~Pac.\ }
\def\prl{Phys. Rev. Lett.\ }
\def\prd{Phys. Rev. D\ }
\def\prc{Phys. Rev. C\ }
\def\prb{Phys. Rev. B\ }
\def\aap{{Astron. Astrophys.}\ }
\def\aa{{Astron. Astrophys.}\ }
\def\apss{{Astrophys. \& Space Sci.}\ }
\def\physrep{{Phys. Repts.}\ }


\onecolumngrid

\clearpage

\printtables
\clearpage
\printfigures

\end{document}